\documentclass[superscriptaddress,aps,letterpaper,10pt,twocolumn,floats,showpacs,amsmath,amsfonts,amssymb,pre,nofootinbib]{revtex4-2}
\usepackage{graphicx}% Include figure files
\usepackage[colorlinks=true,citecolor=blue]{hyperref}
\usepackage[caption=false]{subfig}
\usepackage{pstricks}
\usepackage{pst-node}
\usepackage{amsmath}
\usepackage{amsbsy}
\usepackage{verbatim}
\usepackage{dsfont}
\usepackage{MnSymbol}
\newcommand{\Tr}{\text{Tr}}

\usepackage{amsthm}
\usepackage[T1]{fontenc}

\begin{document}
	
	\title{ Open-system eigenstate thermalization in a noninteracting integrable model}% Force line breaks with \\
	
	\author{Krzysztof Ptaszy\'{n}ski}
   	\email{krzysztof.ptaszynski@ifmpan.poznan.pl}
	\affiliation{Complex Systems and Statistical Mechanics, Department of Physics and Materials Science, University of Luxembourg, L-1511 Luxembourg, Luxembourg}
	\affiliation{Institute of Molecular Physics, Polish Academy of Sciences, Mariana Smoluchowskiego 17, 60-179 Pozna\'{n}, Poland}
	
	\author{Massimiliano Esposito}
	\email{massimiliano.esposito@uni.lu}
	\affiliation{Complex Systems and Statistical Mechanics, Department of Physics and Materials Science, University of Luxembourg, L-1511 Luxembourg, Luxembourg}
	
	\date{\today}
	
	\begin{abstract}
Significant attention has been devoted to the problem of thermalization of observables in isolated quantum setups by individual eigenstates. Here, we address this issue from an open quantum system perspective, examining an isolated setup where a small system (specifically, a single fermionic level) is coupled to a macroscopic fermionic bath. We argue that in such a model, despite its full integrability, the system observables exhibit thermalization when the system-bath setup resides in a typical eigenstate of its Hamiltonian, a phenomenon known as weak eigenstate thermalization. This thermalization occurs unless it is suppressed by localization due to strong coupling. We further show that following the quench of the system Hamiltonian, the system occupancy typically relaxes to the thermal value corresponding to the new Hamiltonian. Finally, we demonstrate that system thermalization also arises when the system is coupled to a bath that has been initialized in a typical eigenstate of its Hamiltonian. Our findings suggest that nonintegrability is not the sole driver of thermalization,  highlighting the need for complementary approaches to fully understand the emergence of statistical mechanics.
	\end{abstract}
	
	\maketitle
	
	\section{Introduction}

Many states of matter occurring in nature can be characterized as thermal equilibrium states, i.e., the expected values of their observables can be characterized using a single or a few intensive thermodynamic parameters (e.g., temperature or chemical potentials of particles). The emergence of thermal behavior has been explained using the concept of statistical ensemble (e.g., microcanonical, canonical, or grand canonical). For example, within the microcanonical picture, the thermal state is described as a statistical mixture of equally probable microstates that have the same energy.

Since the time of Boltzmann, the justification of the statistical assumption that system's microstates are equally probable, based on the principles of classical or quantum mechanics, has been a subject of intense debate~\cite{lebowitz1993boltzmann,lebowitz1999selective,ruelle2004conversations}. For classical dynamical systems, a proposed bridge between mechanics and statistical behavior is the \textit{ergodic hypothesis}. In its strong version, it states that the system's trajectory explores the entire phase space, so that the long-time averages of observables correspond to microcanonical ensemble averages. This property requires that the system is \textit{nonintegrable}, that is, does not have an extensive number of conserved quantities~\cite{dalessio2016quantum,gogolin2016equilibration}.\footnote{However, weakly nonintegrable systems may be not ergodic, as numerically shown for the Fermi--Pasta--Ulam--Tsingou model~\cite{fermi1955studies, berman2005fermi, dauxois2008fermi} and theoretically demonstrated by the Kolmogorov--Arnold--Moser theorem~\cite{kolmogorov1954conservation, arnold1963proof, moser1962invariant}.} In the quantum context, investigations of the origin of statistical mechanics gave rise to the field of \textit{pure-state statistical mechanics}, which postulates and justifies the thermal behavior of individual quantum pure states~\cite{gogolin2016equilibration}. One of the most successful approaches in this area is based on the eigenstate thermalization hypothesis (ETH)~\cite{deutsch1991quantum, srednicki1994chaos, deutsch2018eigenstate,dalessio2016quantum}, which is akin to the classical postulate that the dynamics is ergodic. A strong version of this hypothesis states that observables take thermal values for every eigenstate of the Hamiltonian. ETH has been numerically verified for a wide range of quantum systems far from integrability (for references, see Ref.~\cite{deutsch2018eigenstate}), while it is violated for integrable systems, whose observables are rather described by the so-called generalized Gibbs ensembles~\cite{rigol2007relaxation, cassidy2011generalized, ilievski2015complete}.\footnote{Here, following Ref.~\cite{mori2018thermalization}, by \textit{quantum integrability} we also mean the presence of an extensive number of conserved quantities. However, we note that the concept of integrability is not unequivocally defined for quantum systems~\cite{gogolin2016equilibration,caux2011remarks}.} Furthermore, ETH can be used not only to explain the thermal behavior of static observables but also certain aspects of nonequilibrium dynamics. For example, it justifies thermalization after quench, that is, a situation in which, after a sudden change of the Hamiltonian, the observables relax over time to thermal values for the new Hamiltonian~\cite{polkovnikov2011colloquium}. Furthermore, it can also be used to establish the microscopic origin of open-system thermalization~\cite{fialko2015quantum,odonovan2024quantum}, the second law of thermodynamics and fluctuation theorems~\cite{iyoda2017fluctuation, iyoda2022eigenstate}, or the emergence of nonequilibrium steady states~\cite{xu2022typicality,xu2023emergence,zhang2024emergence}. In fact, ETH has been shown to be necessary to explain certain aspects of thermalization, such as equilibration with thermal baths~\cite{de2015necessity} or universality of dynamics after Hamiltonian quench~\cite{bartsch2017necessity}.

Nevertheless, certain features of thermal behavior can also be observed for integrable systems. Approaches to explain the origin of thermalization without invoking nonintegrability are mostly based on typicality of initial conditions for a system consisting of many degrees of freedom (in the spirit of Boltzmann)~\cite{lebowitz1993macroscopic} or the properties of physically relevant observables (e.g., their typicality or the applicability of coarse-graining). Following the former direction, Chakraborti \textit{et al.}~\cite{chakraborti2022entropy} observed the thermal behavior of a gas of noninteracting classical particles for typical individual microstates randomly sampled from the canonical ensemble. In the latter vein, Khinchin~\cite{khinchin1949mathematical} proposed a weak version of ergodic hypothesis, showing that coarse-grained observables, expressed as sums of many independent degrees of freedom (e.g., coordinates of noninteracting particles), tend to thermalize due to the law of large numbers, without requiring that the dynamics is ergodic at the microscopic level. Mazur and van der Linden~\cite{mazur1963asymptotic} later generalized this result to short-range interacting particles. Building on this idea, Baldovin \textit{et al.}~\cite{baldovin2021statistical} demonstrated thermalization of collective observables in an anharmonic but integrable Toda chain even for highly untypical initial conditions. Later studies~\cite{cocciaglia2022thermalization, baldovin2023ergodic} have even demonstrated thermalization of certain observables in a fully harmonic oscillator chain.

In the quantum context, typicality arguments have been employed in von Neumann's ergodic theorem~\cite{von2010proof, goldstein2010normal}. It states that for typical observables, for every pure state with a narrow energy distribution (i.e., a superposition of energy eigenstates from a microcanonical shell), the time average of the observable corresponds to its microcanonical average. This theorem was later generalized by Reimann~\cite{reimann2015generalization}, who relaxed some of its original assumptions. However, von Neumann's proof has been shown to involve an assumption which is essentially equivalent to ETH~\cite{rigol2012alternatives}. The other concept is canonical typicality~\cite{tasaki1998quantum, gemmer2003distribution, goldstein2006canonical, popescu2006entanglement}, which states that typical pure states with a narrow energy distribution predict the same reduced states of small subsystems as the microcanonical ensemble. This idea has also been generalized to typical pure states with a defined average energy, without imposing restrictions on its distribution~\cite{muller2011concentration}. A related but different concept is \textit{weak eigenstate thermalization hypothesis}, which postulates that subsystem thermalization is also observed for typical (though not all) individual energy eigenstates. This has been proven for translationally invariant lattice models with short-ranged interactions, including integrable ones~\cite{biroli2010effect, mori2016weak, iyoda2017fluctuation}. We note that the applicability of these concepts is often limited compared to ``strong'' ETH. For example, integrable systems obeying weak ETH may not exhibit thermalization after Hamiltonian quench~\cite{biroli2010effect, nandy2016eigenstate}, even for typical initial conditions, due to strong overlap of the initial state with rare nonthermal eigenstates of the final Hamiltonian. Nevertheless, thermalization after quench has been observed for certain integrable models~\cite{dag2018classification}, as well as some nonintegrable systems that violate strong ETH~\cite{mori2017thermalization, harrow2022thermalization}.

Furthermore, certain features of thermalization have been observed even for noninteracting systems described by quadratic Hamiltonians. For example, thermalization of local observables has been observed for typical many-body eigenstates of random noninteracting fermionic~\cite{magan2016random} and bosonic~\cite{cattaneo2024thermalization} Hamiltonians, as well as for translationally invariant models of noninteracting bosons~\cite{biroli2010effect} or fermions~\cite{lai2015entanglement, li2016quantum, nandy2016eigenstate}. However, in the latter case, thermalization can be suppressed by the presence of a disorder that breaks the translational invariance~\cite{li2016quantum}. Still, Ref.~\cite{lydzba2024normal} demonstrated weak eigenstate thermalization of one- and few-body observables for disordered models that exhibit chaos in the single-particle sector. Finally, in a nonequilibrium context, it was shown that in translationally invariant fermionic chains, the charge distribution tends to thermalize at the coarse-grained (macroscopic) level even for highly untypical initial conditions, even though the fine-grained (microscopic) observables do not thermalize~\cite{shiraishi2023nature, tasaki2024macroscopic}. 

Here, we address the problem of pure-state thermalization in integrable and noninteracting models from an open quantum system perspective. By this, we mean that we divide the total closed setup, initialized in an individual eigenstate, into a small system and the large bath, which determines the choice of the considered observables (i.e., they are defined separately for the system and the bath). This approach is common in open quantum system theory~\cite{breuer2002theory, schaller2014open,jakvsic1996model, bach2000return, merkli2001positive, frohlich2004another, merkli2007decoherence, merkli2008resonance, trushechkin2022open}, yet it has rarely been applied in the context of pure-state statistical mechanics. Instead, previous studies of open quantum system thermalization focused on the situation in which either the bath or the system-bath setup was initialized in the thermal state~\cite{jakvsic1996model, bach2000return, merkli2001positive, frohlich2004another, merkli2007decoherence, merkli2008resonance, trushechkin2022open}. As an exception, Usui \textit{et al.}~\cite{usui2023microscopic} recently numerically demonstrated the emergence of nonequilibrium steady states and long-time thermalization of macroscopic bath observables for a fermionic impurity coupled to noninteracting baths initialized in the eigenstates randomly sampled from the grand canonical ensemble, while Purkayastha \textit{et al.}~\cite{purkayastha2024difference} explored the system thermalization after quench in a system-bath setup initialized in typical pure states (superpositions of eigenstates) sampled from the same ensemble. In our study, we examine thermalization in a fully integrable noninteracting resonant level, one of the paradigmatic models in open quantum system theory~\cite{schaller2014open,haug2008quantum}. It consists of a small system---a single fermionic level---coupled to a bath of fermionic levels via a quadratic tunneling Hamiltonian.
We show that the system occupancy exhibits weak eigenstate thermalization, meaning that when the system-bath setup is in a typical eigenstate of its Hamiltonian, the system occupancy tends to have the same value as for the thermal state with the same energy and particle number.  This result extends beyond previously known instances of weak eigenstate thermalization in translationally invariant~\cite{iyoda2017fluctuation,biroli2010effect,nandy2016eigenstate,lai2015entanglement,li2016quantum,dag2018classification}, random~\cite{magan2016random,cattaneo2024thermalization} or chaotic~\cite{lydzba2024normal} models. Specifically, this thermalization occurs when the single-particle state corresponding to the occupied state of the system is strongly delocalized over many single-particle eigenstates of the Hamiltonian. Consequently, thermalization may be suppressed by localization induced by formation of the bound states for a strong system-bath coupling. In addition, no thermalization is observed for occupancies of the bath levels. We further demonstrate that after the quench of the system energy, the system occupancy relaxes to the thermal value for the new Hamiltonian. Finally, we show that thermalization of the system can be induced by the bath initialized in a typical eigenstate of its Hamiltonian.

The paper is organized as follows. In Sec.~\ref{sec:mod} we present details of the considered model and methods used to describe it. In Sec.~\ref{sec:def} we define several concepts and quantities used throughout the paper. In Secs.~\ref{sec:stat} and~\ref{sec:quench} we explore thermalization of static observables and thermalization after quench, respectively. In Sec.~\ref{sec:bath} we investigate the system thermalization induced by typical bath eigenstates. Finally, Sec.~\ref{sec:concl} brings conclusions that follow from our results.
	
	\section{Model and methods} \label{sec:mod}
We consider the noninteracting resonant level, one of the paradigmatic models in open quantum system theory~\cite{schaller2014open,haug2008quantum}. This model is fully integrable, which facilitates the exact study of its equilibrium and nonequilibrium behavior. It is described by the Hamiltonian	
	\begin{align} \label{hamnrl}
		\hat{H}=\epsilon_{0} \hat{c}^\dagger_{0} \hat{c}_{0} +\sum_{k=1}^{K-1} \epsilon_{k} \hat{c}_{k}^\dagger \hat{c}_{k} + \sum_{k=1}^{K-1} \left( t_k \hat{c}^\dagger_{0} \hat{c}_{k} + \text{h.c.} \right),
	\end{align}
where the index $k=0$ corresponds to the system, while $k \in \{1,\ldots,K-1\}$ to the energy levels in the bath. Here $\epsilon_k$ is the level energy, $\hat{c}^\dagger_{k}$ and $\hat{c}_{k}$ are the creation and annihilation operators, $t_k$ is the tunnel coupling between the levels $0$ and $k$, and $K-1$ is the number of energy levels in the reservoir. We further model the density of states in the bath by the boxcar-shaped function, with the bath levels parameterized as $\epsilon_k=-W/2+\Delta_\epsilon(k-1)$, where $\Delta_\epsilon=W/(K-2)$ is the interlevel spacing in the bath, and $W$ is the bath bandwidth. We also use a homogeneous parameterization of the tunnel couplings $t_k=\sqrt{\Gamma/(2\pi \Delta_\epsilon)}$, where $\Gamma$ is the coupling strength to the bath.

This Hamiltonian can be rewritten in the form
		\begin{align} \label{hamnrls}
		\hat{H}=\sum_{k,l} \mathcal{H}_{kl} \hat{c}_k^\dagger \hat{c}_l,
	\end{align}
	where $\mathcal{H}$ is the matrix representing the single-particle sector of the Hamiltonian. It is defined as
	\begin{align} \label{fermhamsp}
		\begin{cases}
			\mathcal{H}_{kk}= \epsilon_k & \text{for} \quad k=0,\ldots,K-1, \\
			\mathcal{H}_{0k}=\mathcal{H}_{k0}=t_k & \text{for} \quad k=1,\ldots,K-1, \\
			\mathcal{H}_{kl}=0 &  \text{otherwise}.
		\end{cases}
	\end{align}

Hamiltonian~\eqref{hamnrl} can be diagonalized to a form
	\begin{align}
		\hat{H}=\sum_{l=0}^{K-1} \omega_l \hat{d}^\dagger_l \hat{d}_l,
	\end{align}
with the mapping between the original and new fermionic operators given as:
	\begin{align} \label{ctod}
		\hat{d}_l=\sum_{k=0}^{K-1} a_{kl}\hat{c}_k, \quad
	\hat{c}_k=\sum_{l=0}^{K-1} a_{kl}^* \hat{d}_l.
	\end{align}
In practice, $\omega_l$ is $l$th eigenvalue of the matrix $\mathcal{H}$, and $a_{kl}^*$ is the $k$th elements of its $l$th normalized eigenvector. We later refer to the levels $l$ as to the \textit{normal modes} of the Hamiltonian. As one can see, the model is highly integrable, since the state of each normal mode is conserved by the Hamiltonian.

The many-particle eigenstates of the Hamiltonian~\eqref{hamnrl} can be expressed as
	\begin{align} \label{eq:manyparteigestates}
	|E_i,N_i \rangle = (\hat{d}_{K-1}^\dagger)^{n_{i,K-1}} \ldots (\hat{d}_{0}^\dagger)^{n_{i,0}} |\varnothing \rangle,
\end{align}
where $|\varnothing \rangle$ is the vacuum state. In particular, the occupied states of individual normal modes $\hat{d}_l^\dagger |\varnothing \rangle$ are called the single-particle eigenstates. Each eigenstate is associated with a unique combination of normal mode occupancies $n_{i,l} \in \{0,1\}$. $E_i =\sum_{l=0}^{K-1} n_{i,l} \omega_l$ denotes the eigenstate energy, and $N_i=\sum_{l=0}^{K-1} n_{i,l}$ is the particle number. We note that for typical many-particle eigenstates, the occupancies of the normal modes are far from being thermal: they can take only values 0 or 1, while thermal occupancies may lie within the range $[0,1]$.

We also consider the case in which, at some moment, the Hamiltonian is quenched from the initial form $\hat{H}$ to the final form $\hat{H}'$. The new Hamiltonian can be analogously diagonalized as
\begin{align}
	\hat{H}'=\sum_{m=0}^{K-1} \nu_m \hat{f}_m^\dagger \hat{f}_m,
\end{align}
with
\begin{align}
	\hat{f}_m = \sum_{k=0}^{K-1} b_{km} \hat{c}_k, \quad
	\hat{c}_k = \sum_{m=0}^{K-1} b_{km}^* \hat{f}_m.
\end{align}

To describe the dynamics of the system under the Hamiltonian $\hat{H}'$, we use the evolution of operators in the Heisenberg picture. Using the expression above, the operators $\hat{c}_k(t)$ evolve as (we take $\hbar=1$)
\begin{align}
	\hat{c}_k(t) =\sum_m b_{km}^* \hat{f}_m e^{-i\nu_m}=\sum_{m,n} b_{nm} b_{km}^* e^{-i \nu_m} \hat{c}_n.
\end{align}
It is also useful to reexpress this formula in terms of operators $\hat{d}_l$. Using Eq.~\eqref{ctod} one obtains
\begin{align} \label{ewop}
	\hat{c}_k(t)=\sum_{l} a_{kl}^*(t) \hat{d}_l,
\end{align}
where
\begin{align} 
a_{kl}(t)=\sum_{n,m}	a_{nl} b_{km} b_{nm}^* e^{i \nu_m}.
\end{align}

Finally, let us understand the amplitudes $a_{kl}$ and $a_{kl}(t)$ as elements of the matrices $\mathbf{a}$ and $\mathbf{a}(t)$, respectively. Then, using spectral decomposition of the matrix exponent, the equation above can be rewritten in a concise form
\begin{align}
\mathbf{a}(t)= e^{i \mathcal{H}'t} \mathbf{a},
\end{align}
which is very convenient for numerical implementation.
	
	\section{Definitions} \label{sec:def}
Let us now define certain concepts and quantities used throughout the paper. First, the microcanonical set $\mathcal{W}$ is defined as a set of energy eigenstates $|E_i,N_i \rangle$ with $E_i \in [E,E+\Delta E]$ and $N_i \in [N,N+\Delta N]$, where $\Delta E$ and $\Delta N$ are small widths of the microcanonical shell. The microcanonical state is defined as an equally weighted mixture of states belonging to the microcanonical set:	
	\begin{align} \label{eq:microcan}
		\rho_\text{mc} \equiv \frac{1}{|\mathcal{W}|} \sum_{i \in \mathcal{W}} |E_i,N_i \rangle \langle E_i,N_i|,
	\end{align}
where $|\mathcal{W}|$ is the cardinality of the microcanonical set, i.e., the number of its elements. We further define the grand canonical state equivalent to the microcanonical state as 
	\begin{align}
	\rho_\text{gc} \equiv \frac{1}{Z} \sum_{i} e^{-\beta(E_i-\mu N_i)} |E_i,N_i \rangle \langle E_i,N_i|,
\end{align}
where $Z=\sum_i e^{-\beta(E_i-\mu N_i)}$ is the partition function. To provide equivalence with the microcanonical state, the inverse temperature $\beta$ and the chemical potential $\mu$ are chosen such that
\begin{align}
\Tr (\rho_\text{gc} \hat{H})=E \quad \text{and} \quad \Tr (\rho_\text{gc} \hat{N})=N,
\end{align}
where $\hat{N} \equiv \sum_{k=0}^{K-1} \hat{c}_k^\dagger \hat{c}_k=\sum_i N_i |E_i,N_i \rangle \langle E_i,N_i|$ is the total particle number operator.

Following Ref.~\cite{mori2018thermalization}, let us now define the \textit{indicator of eigenstate thermalization} (also called \textit{ETH noise}~\cite{dag2018classification, ikeda2013finite}) for the observable $\hat{A}$,
\begin{align} \label{indicator}
	I_\text{mc}(\hat{A}) \equiv \sqrt{\frac{1}{|\mathcal{W}|} \sum_{i \in \mathcal{W}} (\langle \hat{A} \rangle_i-\langle \hat{A} \rangle_{\text{mc}})^2},
\end{align}
where $\langle \hat{A} \rangle_i = \langle E_i,N_i|\hat{A}|E_i,N_i \rangle$ is the expected value of the observable $\hat{A}$ for an individual eigenstate $|E_i,N_i \rangle$, and $\langle \hat{A} \rangle_\text{mc}=\Tr (\rho_\text{mc} \hat{A})$ is 
the microcanonical average of this observable. This quantity measures how, typically, the expected value of the observable $\hat{A}$ evaluated for an individual eigenstate differs from its thermal value. In other words, it evaluates how well typical eigenstates belonging to the microcanonical set thermalize the observable $\hat{A}$. Thermalization is indicated by the vanishing of $I_\text{mc}(\hat{A})$ in the thermodynamic limit. In particular, in nonintegrable models obeying strong ETH (thermalization for all eigenstates) one observes a polynomial decay of this quantity with dimension of the Hilbert space of the system (and thus often exponential with system size)~\cite{mori2018thermalization, steinigeweg2013eigenstate, steinigeweg2014pushing, beugeling2014finite}, while in integrable models obeying weak ETH (thermalization for typical eigenstates) the decay is slower (often polynomial with system size~\cite{mori2018thermalization, alba2015eigenstate, biroli2010effect, ikeda2013finite}).

\section{Thermalization of static observables} \label{sec:stat}
\subsection{Theoretical arguments} \label{subsec:stattheor}
We now analyze whether and which observables of the considered system are thermalized by single eigenstates $|E_i,N_i \rangle$. In particular, we focus on the local occupancies of levels $k$ in the original basis. They are denoted as $\langle \hat{p}_k \rangle$, where $\hat{p}_k=\hat{c}_k^\dagger \hat{c}_k$. We note that, due to the parity superselection rule that prohibits coherent superpositions of the empty and occupied state of a single level~\cite{wick1952intrinsic,szalay2021fermionic}, this observable fully defines the reduced state of each level $k$, namely, its reduced density matrix takes the form $\rho_k=\text{diag}(1-\langle \hat{p}_k \rangle,\langle \hat{p}_k \rangle)$ in the basis of the empty and occupied states. Consequently, any other local observable for the level $k$ is also fully determined by $\langle \hat{p}_k \rangle$.

To determine this quantity, we use the fact that, for both single eigenstates and for the microcanonical or grand canonical states, the density matrix of the system is diagonal in the basis diagonalizing the Hamiltonian $\hat{H}$. As a consequence, occupancies $\langle \hat{p}_k \rangle$ can be expressed in terms of occupancies of normal modes:
\begin{align} \label{popstat}
\langle \hat{p}_k \rangle= \sum_{l=0}^{K-1} |a_{kl}|^2 \langle \hat{n}_l \rangle,
\end{align}
where $\hat{n}_l = \hat{d}_l^\dagger \hat{d}_l$. Let us now define the observable $\hat{\pi}_k \equiv \sum_{l=0}^{K-1} |a_{kl}|^2 \hat{n}_l$, which commutes with the system Hamiltonian so that $ \hat{\pi}_k |E_i,N_i \rangle=\langle \hat{p}_k \rangle_i |E_i,N_i \rangle$ (i.e., the energy eigenstates $|E_i,N_i\rangle$ are also the eigenstates of $\hat{\pi}_k$). Using Eq.~\eqref{indicator}, the indicator of eigenstate thermalization of the observable $\hat{p}_k$ can be calculated as
\begin{align}
I_\text{mc}(\hat{p}_k)=\sqrt{\text{Var}_\text{mc}(\hat{\pi}_k)},
\end{align}
where 
\begin{align}
\label{vmic}
\text{Var}_\text{mc}(\hat{\pi}_k) \equiv \Tr[\rho_\text{mc} (\hat{\pi}_k-\langle \hat{\pi}_k \rangle_\text{mc})^2]
\end{align} 
is the microcanonical variance of the observable $\hat{\pi}_k$. This can be easily derived by inserting the definition of the microcanonical state [Eq.~\eqref{eq:microcan}] to the expression above, and later using  $\hat{\pi}_k |E_i,N_i \rangle=\langle \hat{p}_k \rangle_i |E_i,N_i \rangle$, $\langle \hat{\pi}_k \rangle_\text{mc}=\langle \hat{p}_k \rangle_\text{mc}$.

We now note that, for the model considered, the overlaps $|a_{kl}|^2$ are significant only for those normal modes $l$ whose energies $\omega_l$ are close to $\epsilon_k$ (i.e., the normal modes resonantly coupled with the level $k$). We now assume that the number of such resonant normal modes is much lower than the number of all modes $K$, but may still be extensive with $K$. As later discussed in Sec.~\ref{subsec:local}, this may be expected when the coupling strength $\Gamma$ is much smaller than the bandwidth $W$. Consequently, the observable $\hat{\pi}_k$ is effectively confined only to the subsystem corresponding to the set of resonant normal modes, which is much smaller than the set of all normal modes. Then, according to the principle of ensemble equivalence, when the total system-bath setup is in the microcanonical state, the state of such small subsystem corresponds to the grand canonical state. Consequently, we can approximate the microcanonical variance $\text{Var}_\text{mc}(\hat{\pi}_k)$ by its grand canonical equivalent $\text{Var}_\text{gc}(\hat{\pi}_k)$, that is  defined by replacing $\rho_\text{mc}$ with $\rho_\text{gc}$ in Eq.~\eqref{vmic}. We get 
\begin{align} \label{eq:enseqapp}
	&I_\text{mc}(\hat{p}_k)\approx \sqrt{\text{Var}_\text{gc} (\hat{\pi}_k)} = \sqrt{\sum_{l} |a_{kl}|^4 \langle \hat{n}_l \rangle_\text{gc} (1-\langle \hat{n}_l \rangle_\text{gc})},
\end{align}
where $\langle \hat{A} \rangle_\text{gc} =\Tr[\rho_\text{gc} \hat{A}]$ denotes the expected value in the grand canonical state. Since $\langle \hat{n}_l \rangle_\text{gc} \in [0,1]$, we can approximately bound the indicator $I_\text{mc}(\hat{p}_k)$ as
\begin{align} \label{iprbound}
	I_\text{mc}(\hat{p}_k) \lessapprox \frac{1}{2}\sqrt{\text{IPR}_k},
\end{align}
where $\text{IPR}_k$ is the inverse participation ratio for level $k$:
\begin{align}
\text{IPR}_k \equiv \sum_{l=0}^{K-1} |a_{kl}|^4.
\end{align}
This quantity is a standard measure of the delocalization of single-particle states $\hat{c}_k^\dagger|\varnothing \rangle$ over single-particle eigenstates $\hat{d}_l^\dagger|\varnothing \rangle$~\cite{kramer1993localization}. By virtue of Eq.~\eqref{popstat}, it also describes the delocalization of the occupancy $\langle \hat{p}_k \rangle$ over the occupancies of the normal modes $\langle \hat{n}_l \rangle$ (for simplicity, we will call it delocalization over the normal modes): When $\text{IPR}_k$ is close to $1$, it is strongly localized in a single normal mode. In contrast, it is delocalized when it goes to 0 in the thermodynamic limit. Consequently, Eq.~\eqref{iprbound} implies that occupancies $\langle \hat{p}_k \rangle_i$ are thermalized by typical eigenstates of the system-bath setup when they are strongly delocalized. This reminds us of the result of Khinchin, obtained in the context of classical statistical mechanics: Observables, which can be expressed as a sum of many independent degrees of freedom, tend to thermalize due to the law of large numbers, independently of the details of microscopic dynamics~\cite{khinchin1949mathematical}.\footnote{We note that this idea was also used in the manuscript submitted after our work to rationalize thermalization of random observables, and local observables for random Hamiltonians, in bosonic noninteracting systems~\cite{cattaneo2024thermalization}.} We also note that the connection between thermalization or equilibration and the inverse participation ratio (or its inverse, known as the effective dimension) has already been recognized in the literature~\cite{linden2009quantum, short2011equilibration, short2012quantum, gogolin2016equilibration, farelly2017thermalization}.

We emphasize, however, that even for delocalized occupancies $\langle \hat{p}_k \rangle$ one can find very untypical eigenstates in the microcanonical set with strongly nonthermal values of $\langle \hat{p}_k \rangle_i$. For example, there may exist eigenstates in which all normal modes $l$ that are resonant with level $k$ have populations of either 0 or 1. Thus, in general, the distribution of expected values $\langle \hat{p}_k \rangle_i$ has support that extends over the nearly the entire [0,1] interval (i.e., $\max_i \langle \hat{p}_k \rangle_i-\min_i \langle \hat{p}_k \rangle_i$ is close to 1), and therefore some expected values can strongly deviate from the thermal value. However, for strongly delocalized occupancies $\langle \hat{p}_k \rangle$, eigenstates with nonthermal expected values are very rare. To formalize this statement, let us quantify the fraction of eigenstates for which the absolute deviation from the thermal value $|\langle \hat{p}_k \rangle_i-\langle \hat{p}_k \rangle_\text{mc}|$ is greater than some number $\xi$. According to the law of large numbers, this is equal to the probability of such an occurrence for eigenstates randomly drawn from the microcanonical set, which is denoted $P(|\langle \hat{p}_k \rangle_i-\langle \hat{p}_k \rangle_\text{mc}| \geq \xi)$. Using Eq.~\eqref{indicator}, we can further identify the indicator $I_\text{mc} (\hat{p}_k)$ as the standard deviation of the probability distribution of expected values $\langle \hat{p}_k \rangle_i$ for such randomly drawn eigenstates. Then, the probability of deviation from the thermal value can be bounded by Chebyshev's inequality~\cite{mori2018thermalization} 
\begin{align}
P(|\langle \hat{p}_k \rangle_i-\langle \hat{p}_k \rangle_\text{mc}| \geq \xi) \leq \frac{I_\text{mc}(\hat{p}_k) ^2}{\xi^2} \lessapprox \frac{\text{IPR}_k}{4 \xi^2}.
\end{align}
This implies that for strongly delocalized occupancies $\langle \hat{p}_k \rangle$ (with $\text{IPR}_k \rightarrow 0$) the fraction of eigenstates for which $\langle \hat{p}_k \rangle_i$ deviates from the thermal value by any positive $\xi$ goes to 0. This situation is referred to as weak eigenstate thermalization (to distinguish it from strong eigenstate thermalization, which occurs for all eigenstates).

\subsection{Thermalization for typical eigenstates from the grand canonical ensemble} \label{subsec:grand}
Our paper focuses on the physically motivated question of whether, for typical eigenstates, the observables have the same expected values as for the thermal states with the same energy and particle number. This is equivalent to asking whether the observables thermalize to the microcanonical value for typical eigenstates randomly drawn from the microcanonical ensemble with a defined energy and particle number. An alternative question is whether the observables thermalize to the grand canonical value for typical eigenstates that are randomly drawn from the grand canonical ensemble with a defined inverse temperature $\beta$ and chemical potential $\mu$ (i.e., the eigenstate $|E_i,N_i\rangle$ is drawn with a probability $e^{-\beta(E_i-\mu N_i)}/Z$). In fact, such thermalization has been recently numerically demonstrated for classical~\cite{chakraborti2022entropy} and quantum~\cite{usui2023microscopic} systems. While this question is somewhat less intuitive than the former one, it is much easier to handle mathematically. Thermalization in the discussed sense is witnessed by the vanishing of the indicator 
\begin{align} \label{indicatorgrand}
	I_\text{gc}(\hat{p}_k) \equiv \sqrt{\sum_{i} \frac{e^{-\beta(E_i-\mu N_i)}}{Z}  (\langle \hat{p}_k \rangle_i-\langle \hat{p}_k \rangle_{\text{gc}})^2},
\end{align}
which, analogously to $I_\text{mc}(\hat{p}_k)$, quantifies the standard deviations of occupancies for individual eigenstates $\langle \hat{p}_k \rangle_i$  from the grand canonical average $\langle \hat{p}_k \rangle_{\text{gc}}$. Following the steps from the previous subsection, it can be bounded as
\begin{align} \label{eq:iprboundgrand}
I_\text{gc}(\hat{p}_k) =\sqrt{\text{Var}_\text{gc} (\hat{\pi}_k)} \leq \frac{1}{2} \sqrt{\text{IPR}_k}.
\end{align}
Compared to Eq.~\eqref{iprbound}, this expression is completely exact, as it does not rely on the assumption of applicability of the ensemble equivalence. Thus, delocalization of the occupancy $\langle \hat{p}_k \rangle$ (i.e., vanishing of $\text{IPR}_k$ in the thermodynamic limit) is a sufficient condition of its thermalization for typical eigenstates randomly drawn from the grand canonical ensemble.

\subsection{Localization} \label{subsec:local}
%%%%%%%%%%%%%%%%%%%%%%%%%%%%%%%%%%%%%%%%%%%%%%%%%%%%%%%%%%%%%%%%%%%%
\begin{figure}
	\centering
	\includegraphics[width=0.9\linewidth]{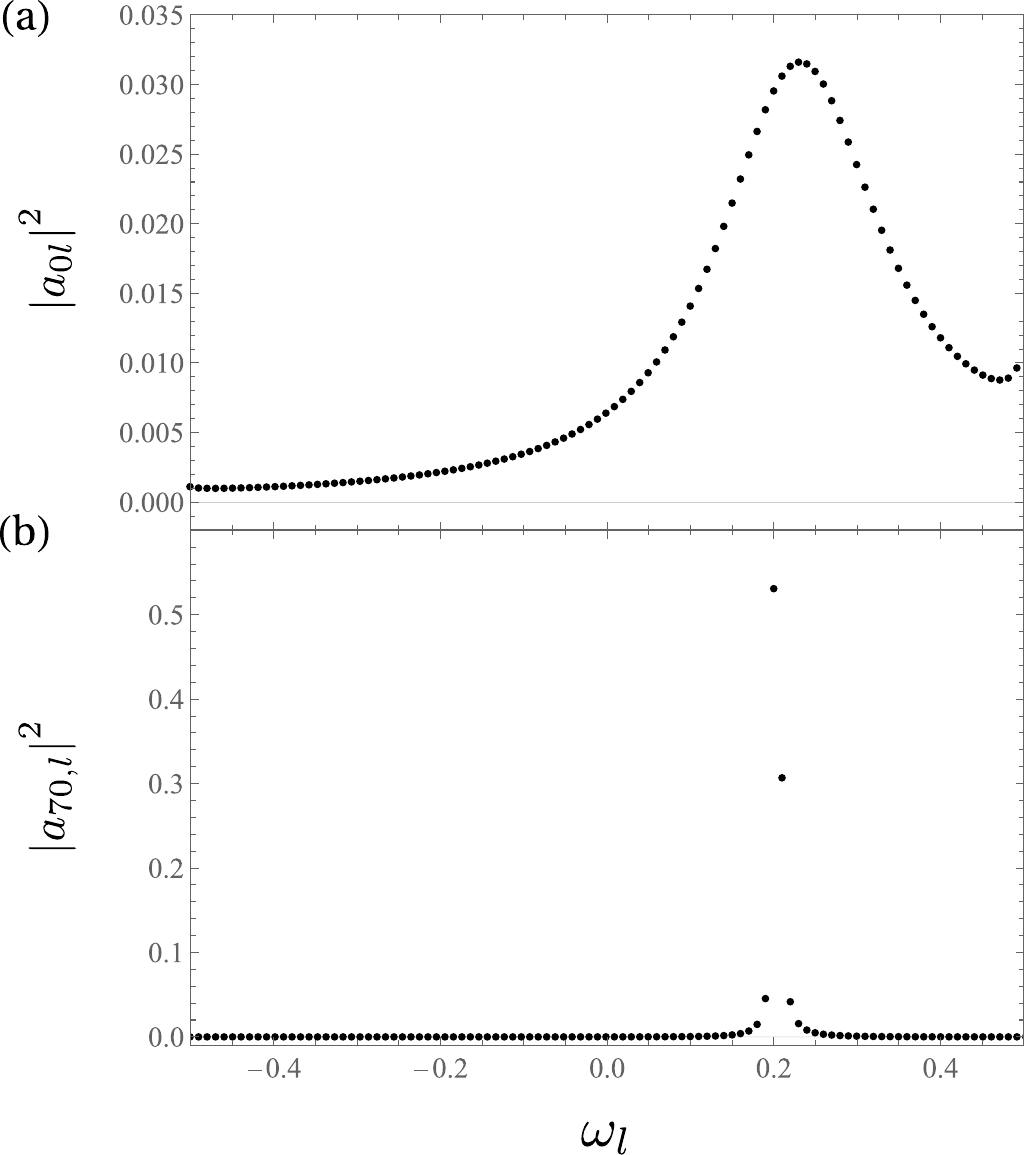}
	\caption{The overlaps $|a_{kl}|^2$ as a function of the normal mode energy $\omega_l$ for the system level $k=0$ (a) and the bath level $k=140$ (b). Parameters: $K=200$, $\epsilon_0=0.2W$, $\Gamma=0.1W$.}
	\label{fig:amplitudes}
\end{figure}
%%%%%%%%%%%%%%%%%%%%%%%%%%%%%%%%%%%%%%%%%%%%%%%%%%%%%%%%%%%%%%%%%%%%
Let us now analyze the localization properties of occupancies $\langle \hat{p}_k \rangle$. First, we analyze the distribution of the overlaps $|a_{0k}|^2$ which quantify contributions of different normal modes to the system occupancy. This is plotted in Fig.~\ref{fig:amplitudes}~(a) for $K=200$. One may observe that the occupancy $\langle \hat{p}_0 \rangle$ is delocalized over many normal modes with energies $\omega_l$ close to $\epsilon_0$. This is because the system is strongly affected by the interaction with many bath levels (provided that the coupling strength $\Gamma$ is neither too weak nor too strong; see a discussion below). At the same time, the overlaps are only significant for $\omega_l$ sufficiently close to $\epsilon_0$ (i.e., for normal modes resonant with the system), in the region of the width $\Gamma$, and are suppressed for $\omega_l$ far from $\epsilon_0$. In fact, in quantum transport theory, $\Gamma$ is treated as an effective broadening of the system energy~\cite{datta2005quantum}. This justifies the application of ensemble equivalence in Eq.~\eqref{eq:enseqapp} provided that the coupling strength $\Gamma$ is sufficiently small compared to the bandwidth $W$. However, the number of significant overlaps is then still extensive with the setup size as $\propto \Gamma K/W$. According to our theoretical reasoning, this leads to thermalization of the system occupancy $\langle \hat{p}_0 \rangle$; we will demonstrate this numerically in Sec.~\ref{subsec:thermstatnum}.

For comparison, let us now consider the localization properties of the occupancies $\langle \hat{p}_k \rangle$ of the bath levels (with $k>0$). Specifically, in Fig.~\ref{fig:amplitudes}~(b) we present the overlaps $|a_{kl}|^2$ for the level $k=140$, that is, the bath level with the energy $\epsilon_k$ closest to $\epsilon_0$. As can be seen, the occupancy $\langle \hat{p}_{140} \rangle$ is strongly localized in just a pair of normal modes with the energies $\omega_l$ closest to $\epsilon_{140}$. This is reasonable, since the bath is only weakly perturbed by a coupling to a small system. As will be shown later, this results in the lack of thermalization of the bath level occupancies.

%%%%%%%%%%%%%%%%%%%%%%%%%%%%%%%%%%%%%%%%%%%%%%%%%%%%%%%%%%%%%%%%%%%%
\begin{figure}
	\centering
	\includegraphics[width=0.9\linewidth]{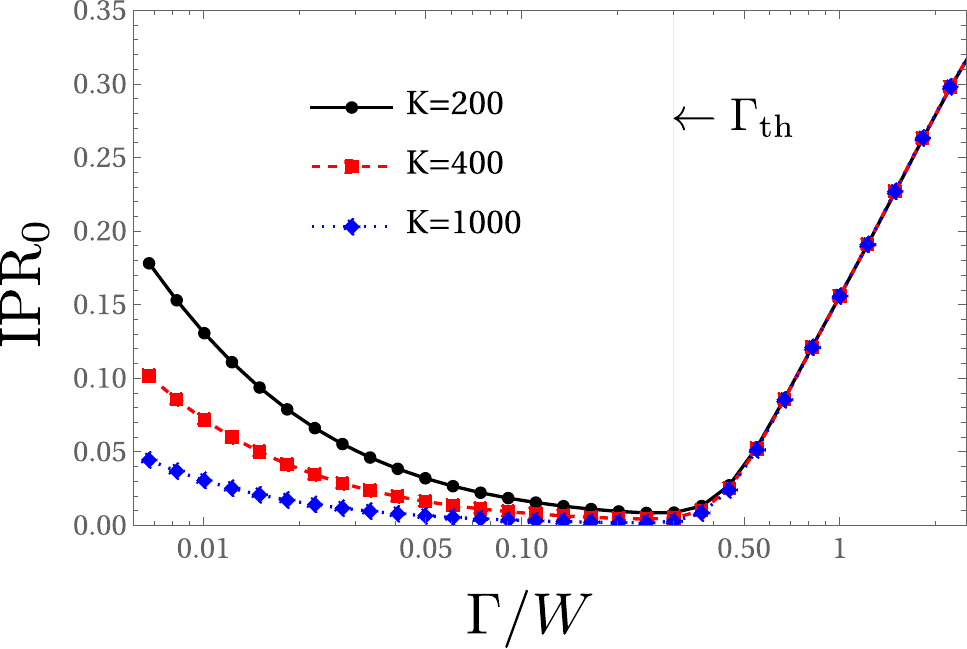}
	\caption{The inverse participation ratio $\text{IPR}_0$ as a function of the coupling strength $\Gamma$ for different number of levels $K$. Parameters: $\epsilon_0=0.2W$.}
	\label{fig:ipr}
\end{figure}
%%%%%%%%%%%%%%%%%%%%%%%%%%%%%%%%%%%%%%%%%%%%%%%%%%%%%%%%%%%%%%%%%%%%
We now focus on the localization properties of the system occupancy by analyzing the inverse participation ratio $\text{IPR}_{0}$. In Fig.~\ref{fig:ipr} we plot this quantity (in the log-linear scale) as a function of the coupling strength $\Gamma$ for different numbers of levels $K$. Interestingly, it exhibits a qualitatively different behavior for $\Gamma$ below and above the localization threshold $\Gamma_\text{th}$. The numerical results suggest that this threshold corresponds to the distance of the system energy from the band edge: $\Gamma_\text{th} \approx W/2-|\epsilon_0|$. Below the threshold, the inverse participation ratio decreases with increasing $\Gamma$ or $K$. This may be explained as follows: As discussed above, the overlaps $|a_{0l}|^2$ are significant only for those levels $l$, whose energies are resonant with the system, that is, for which the difference $|\omega_l-\epsilon_0|$ is of the order of the coupling strength $\Gamma$ (which corresponds to the level broadening due to the system-bath coupling). The number of those levels is proportional to $\Gamma/\Delta_\epsilon \propto \Gamma K$, where, to recall, $\Delta_\epsilon=W/(K-2)$ is the interlevel spacing in the bath. At the same time, the overlaps are normalized as $\sum_{l=0}^{K-1} |a_{kl}|^2=1$. Thus, individual overlaps $|a_{0l}|^2$ scale proportionally to $1/(\Gamma K)$, and their squares $|a_{0l}|^4$ scale as $1/(\Gamma K)^2$. Consequently, the inverse participation ratio $\text{IPR}_0 = \sum_{l=0}^{K-1} |a_{0l}|^4$ scales as $(\Gamma K)/(\Gamma K)^2=1/(\Gamma K)$ (proportionally to the number of significant elements $|a_{0l}|^4$ and inversely proportionally to their values). Thus, the inverse participation ratio tends to vanish in the thermodynamic limit $K \rightarrow \infty$, regardless of the coupling strength $\Gamma$. However, for a finite setup size, the coupling $\Gamma$ must be sufficiently strong to observe delocalization and thus thermalization; a similar observation was made previously for open systems described by random matrix Hamiltonians~\cite{esposito2003spin}.

In contrast, above $\Gamma_\text{th}$ the inverse participation ratio becomes size-independent and grows with $\Gamma$ (i.e., the system occupancy becomes increasingly localized). This is a result of the formation of bound states, i.e., single-particle eigenstates $\hat{d}_l^\dagger|\varnothing \rangle$ strongly localized in the system~\cite{cai2014threshold, xiong2015non, yang2015master, jussiau2019signature}. Therefore, it may be inferred that in the thermodynamic limit $K \rightarrow \infty$ the system exhibits a localization phase transition, with a delocalized phase ($\text{IPR}_0=0$) for $\Gamma<\Gamma_\text{th}$ and a localized phase ($\text{IPR}_0>0$) for $\Gamma>\Gamma_\text{th}$. Consequently, according to our previous theoretical arguments (Sec.~\ref{subsec:stattheor}), when the Hamiltonian parameters correspond to the delocalized phase, the system occupancy $\langle \hat{p}_0 \rangle$ should thermalize in the thermodynamic limit for typical eigenstates randomly drawn from the microcanonical ensemble. Furthermore, it is guaranteed to thermalize for typical eigenstates drawn from the grand canonical ensemble (Sec.~\ref{subsec:grand}).

\subsection{Thermalization: numerical results} \label{subsec:thermstatnum}
Let us now confirm the validity of our reasoning with numerical simulations. Specifically, in our simulations we want to emulate the selection of eigenstates from the microcanonical ensemble (the scenario considered in Sec.~\ref{subsec:stattheor})\footnote{We note that simulations are much simpler in the case of sampling from the grand canonical ensemble, discussed in Sec.~\ref{subsec:grand}. Then, sampling can be performed by randomly generating the populations of normal modes with a probability given by Fermi-Dirac distribution~\cite{usui2023microscopic}. However, in this case, numerical confirmation is actually not needed, as thermalization of delocalized occupancies is rigorously proven by Eq.~\eqref{eq:iprboundgrand}.}. A most natural way to do this would be to directly determine the indicator $I_\text{mc}(\hat{p}_k)$ for a narrow energy window corresponding to the microcanonical shell. However, this is not feasible because the dimension of the Hilbert space of the total system-bath setup (i.e., the number of all many-particle eigenstates) grows exponentially with the number of energy levels as $2^K$. For example, for $K=300$ this dimension is of the order $10^{90}$, $10^{10}$ times larger than the number of atoms in the observed universe. Thus, it is not possible to find all the eigenstates belonging to the microcanonical shell. Therefore, we use another approach. First, we randomly generate eigenstates $|E_i,N_i \rangle$ with a specified particle number $N_i=N$. By virtue of Eq.~\eqref{eq:manyparteigestates}, this corresponds to the generation of a random sequence of normal mode occupancies $n_{i,l} \in \{0, 1\}$ fulfilling $\sum_{l=0}^{K-1} n_{i,l}=N$. Then, for each eigenstate, we determine the reference grand canonical state $\rho_{\text{gc},i}=Z^{-1} \exp[-\beta_i(\hat{H}-\mu_i \hat{N})]$ having the same energy $E_i$ and the particle number $N$. The inverse temperature and chemical potential of the reference state are determined by solving the equations
\begin{align} \label{eq:refen}
E_i &= \sum_{l=0}^{K-1} \omega_l f[\beta_i (\omega_l-\mu_i)], \\ \label{eq:refnum}
N &=\sum_{l=0}^{K-1} f[\beta_i (\omega_l-\mu_i)],
\end{align}
where $f(x)=1/[1+\exp(x)]$ is the Fermi-Dirac distribution. We further focus on the half-filled case with $N=K/2$. 

To emulate the sampling of eigenstates from the microcanonical ensemble, we then select only those eigenstates whose reference temperatures $T_i=1/\beta_i$ (with $k_B=1$) belong to the interval $[T_\text{mid}-\Delta T/2,T_\text{mid}+\Delta T/2]$\footnote{This is equivalent to selecting eigenstates with energies $E_i$ belonging to certain energy interval. $T_i$ acts here as an intensive parameter that parameterizes the state energy.}. Ideally, to emulate the microcanonical shell well, $\Delta T$ should be very small. Unfortunately, in order to make the sampling of eigenstates feasible, it cannot be too small. Indeed, the random generation of a sufficient number of eigenstates with $T_i \in [T_\text{mid}-\Delta T/2,T_\text{mid}+\Delta T/2]$ is the main computational bottleneck in our simulations, which limits our study to relatively small setup sizes $K \leq 300$. This is because the fraction of the corresponding eigenstates decreases with the setup size for a finite $T_\text{mid}$, as most eigenstates tend to have very high reference temperatures $T_i$. Furthermore, to demonstrate the thermalization, in the next paragraph we analyze the dependence of the system occupancy $\langle \hat{p}_0 \rangle_i$ on the system energy $\epsilon_0$. To show this dependence, the midpoint of the interval $T_\text{mid}$ cannot be too large; otherwise, the system tends to trivially thermalize to a fully mixed state with $\langle \hat{p}_0 \rangle_i \approx N/K=1/2$, independent of the system energy. To meet the above conditions, in our simulations we take $T_\text{mid}=0.45W$and $\Delta T=0.1W$. For such parameters, the  fraction of eigenstates corresponding to the reference temperature window (i.e., with $T_i \in [T_\text{mid}-\Delta T/2,T_\text{mid}+\Delta T/2]$) is still very small, being of the order $10^{-6}$ for $K=300$. As a consequence, we consider a relatively small sample of $M=100$ eigenstates, which still required us to randomly generate about $10^8$ random eigenstates.

%%%%%%%%%%%%%%%%%%%%%%%%%%%%%%%%%%%%%%%%%%%%%%%%%%%%%%%%%%%%%%%%%%%%
\begin{figure}
	\centering
	\includegraphics[width=0.9\linewidth]{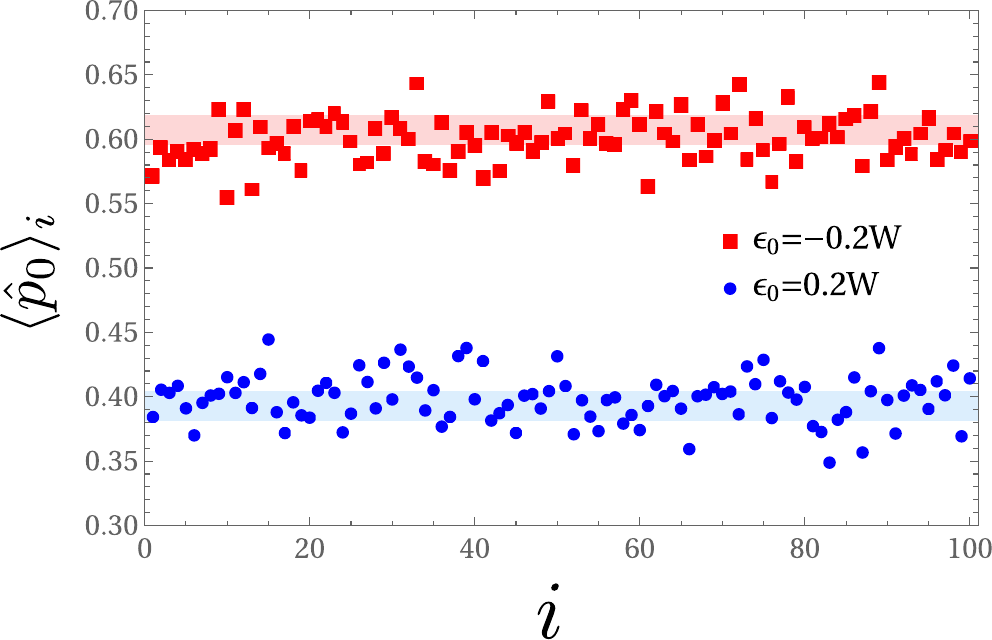}
	\caption{The system occupancies $\langle \hat{p}_0 \rangle_i$ for $M=100$ individual random eigenstates $|E_i,N_i \rangle$ for $\epsilon_0=0.2W$ (blue dots) and $\epsilon_0=-0.2W$ (red squares). The blue- and red-shaded regions denote the range of allowed equilibrium occupancies for $\epsilon_0=0.2W$ and $\epsilon_0=-0.2W$, respectively. Parameters: $\Gamma=0.2W$, $K=300$.}
	\label{fig:popstat}
\end{figure}
%%%%%%%%%%%%%%%%%%%%%%%%%%%%%%%%%%%%%%%%%%%%%%%%%%%%%%%%%%%%%%%%%%%%
In Fig.~\ref{fig:popstat} we present the system occupancies $\langle \hat{p}_0 \rangle_i$ calculated for $M=100$ randomly generated eigenstates $|E_i,N_i \rangle$ for $K=300$, $\Gamma=0.2W$ corresponding to the delocalized phase, and two different system level energies: $\epsilon_0=0.2W$ (blue dots) and the $\epsilon_0=-0.2W$ (red squares). The shaded regions denote the range of allowed equilibrium occupancies of the system for the thermal states with the same mean particle number as the considered eigenstates (i.e., $\langle \hat{N}\rangle=N$) and the temperatures belonging to the considered reference temperature window  $[T_\text{mid}-\Delta T/2,T_\text{mid}+\Delta T/2$]. The blue and red shading of the regions corresponds to the system energies $\epsilon_0=0.2W$ and $\epsilon_0=-0.2W$, respectively. As one can observe, for a given system energy $\epsilon_0$, the system occupancies for individual eigenstates $\langle \hat{p}_0 \rangle_i$ tend to be distributed in or near the corresponding shaded region (i.e., the blue dots corresponding to $\epsilon_0=0.2W$ are distributed near the blue-shaded region, and the red squares corresponding to $\epsilon=-0.2W$ are distributed near the red-shaded region). This witnesses thermalization of the system occupancy.

%%%%%%%%%%%%%%%%%%%%%%%%%%%%%%%%%%%%%%%%%%%%%%%%%%%%%%%%%%%%%%%%%%%%
\begin{figure}
	\centering
	\includegraphics[width=0.9\linewidth]{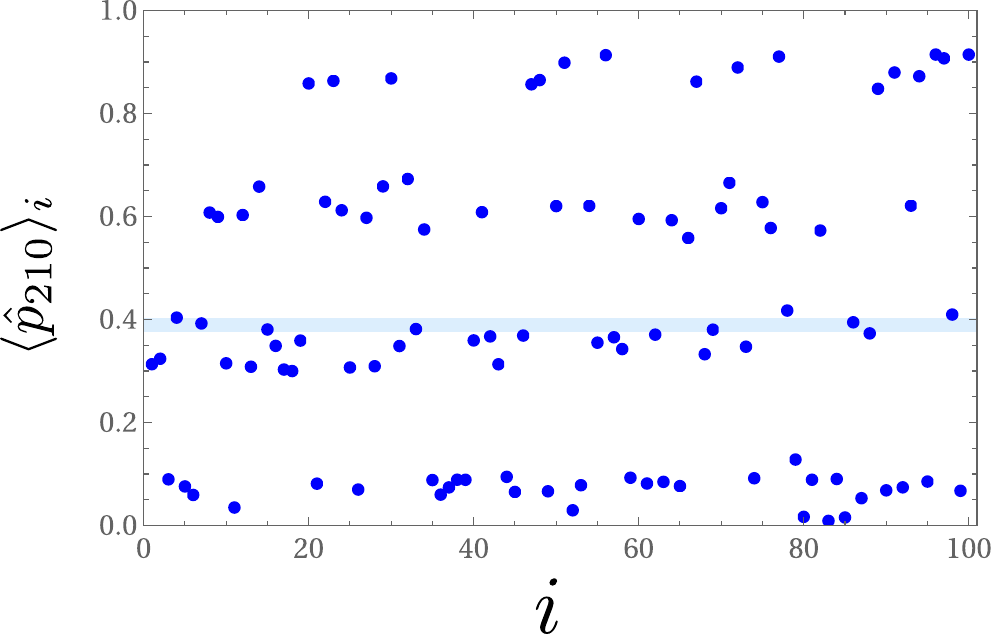}
	\caption{The occupancies of the bath level $k=210$ for $M=100$ individual random eigenstates $|E_i,N_i \rangle$ (blue dots). The blue-shaded region denotes the range of allowed equilibrium occupancies. Parameters: $\epsilon_0=0.2W$, $\Gamma=0.2W$, $K=300$.}
	\label{fig:popstatenv}
\end{figure}
%%%%%%%%%%%%%%%%%%%%%%%%%%%%%%%%%%%%%%%%%%%%%%%%%%%%%%%%%%%%%%%%%%%%
As already suggested by the results on localization properties, a very different behavior is observed for occupancies of the bath levels. This is illustrated in Fig.~\ref{fig:popstatenv} for $\epsilon_0=0.2W$ and the level $k=210$, whose energy $\epsilon_k$ is closest to $\epsilon_0$. As shown, now the level occupancies tend to be distributed around four different bands rather than around the blue-shaded region corresponding to the range of allowed equilibrium occupancies. This implies that, in contrast to the system, the local observables of the bath do not thermalize.

%%%%%%%%%%%%%%%%%%%%%%%%%%%%%%%%%%%%%%%%%%%%%%%%%%%%%%%%%%%%%%%%%%%%
\begin{figure}
	\centering
	\includegraphics[width=0.9\linewidth]{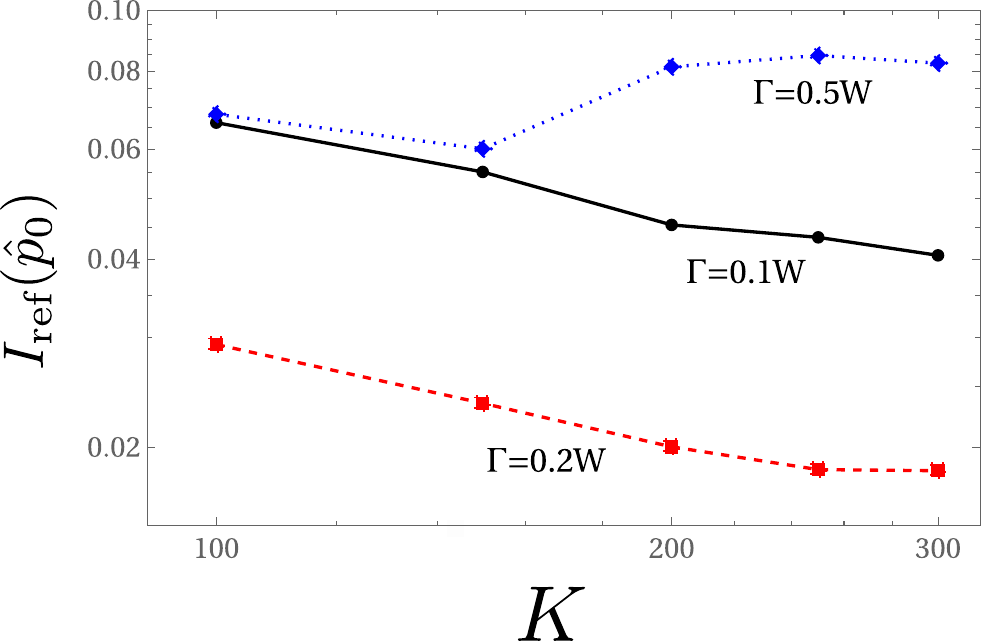}
	\caption{Scaling of the indicator of eigenstate thermalization $I_\text{ref}(\hat{p}_0)$ with the number of levels $K$ for different values of the coupling strength $\Gamma$, evaluated for $M=100$ randomly generated eigenstates. The results are represented by dots, and lines are added for eye guidance. Parameter: $\epsilon_0=0.2 W$.}
	\label{fig:popstatscaling}
\end{figure}
%%%%%%%%%%%%%%%%%%%%%%%%%%%%%%%%%%%%%%%%%%%%%%%%%%%%%%%%%%%%%%%%%%%%
To illustrate the system thermalization further, we calculate a modified indicator of eigenstate thermalization defined as
\begin{align} \label{refindicator}
	I_\text{ref}(\hat{p}_0) \equiv \sqrt{\frac{1}{M} \sum_{i=1}^M (\langle \hat{p}_0 \rangle_i-\langle \hat{p}_0\rangle_{\text{gc},i})^2},
\end{align}
which measures the deviation of the system occupancies evaluated for individual eigenstates, $\langle \hat{p}_0 \rangle_i$, from the occupancies evaluated for the reference thermal states,
\begin{align}
\langle \hat{p}_0 \rangle_{\text{gc},i} \equiv \Tr(\rho_{\text{gc},i} \hat{p}_0 )=\sum_{l=0}^{K-1} |a_{0l}|^2 f[\beta_i (\omega_l-\mu_i)].
\end{align} 
To recall, $M=100$ is here the size of the sample of eigenstates. This quantity well approximates the previously defined indicator $I_\text{mc}(\hat{p}_0)$ provided that the width of the considered reference temperature interval $\Delta T$ is sufficiently small, and $K$ is sufficiently large, such that $\langle \hat{p}_0\rangle_{\text{gc},i}$ and $\langle \hat{p}_0\rangle_{\text{mc}}$ coincide due to ensemble equivalence. In Fig.~\ref{fig:popstatscaling} we present the scaling of $I_\text{ref}(\hat{p}_0)$ with $K$ (in the log-log scale) for different values of the coupling strength $\Gamma$. For $\Gamma=0.1W$ and $\Gamma=0.2W$, corresponding to the delocalized phase, the indicator decreases with $K$ approximately as $1/\sqrt{K}$. Using Eq.~\eqref{iprbound} and the above-discussed relation between $I_\text{mc}(\hat{p}_0)$ and $I_\text{ref}(\hat{p}_0)$, this may be related to the $1/K$ scaling of $\text{IPR}_0$, discussed in Sec.~\ref{subsec:local}. Some deviations from this behavior are related to the finite size of the sample $M=100$. This implies that the system occupancy tends to thermalize in the thermodynamic limit $K \rightarrow \infty$. We note that the same scaling ($1/\sqrt{K}$) of the deviation of delocalized observables from the thermal value has previously been demonstrated by Khinchin in the classical context~\cite{khinchin1949mathematical}. It is also common in translationally invariant~\cite{biroli2010effect,alba2015eigenstate} or chaotic quadratic~\cite{lydzba2024normal} systems that exhibit weak eigenstate thermalization. This can be related to a similar origin of these phenomena from the vanishing of certain types of fluctuations in the microcanonical ensemble (e.g., fluctuations of the delocalized observable $\hat{\pi}_k$ in our case or fluctuations of intensive observables for translationally invariant systems~\cite{biroli2010effect}), which in turn has a common root in the law of large numbers (see a similar scaling of standard deviation of a normalized sum of $K$ independent random variables).
 
In contrast, for $\Gamma=0.5 W$, corresponding to the localized phase, the indicator tends to saturate at a constant value independent of $K$, implying the absence of thermalization. We note that a similar suppression of thermalization by localization was previously observed for noninteracting fermionic lattice models~\cite{li2016quantum,lydzba2024normal}, as well as interacting nonintegrable systems, where it is referred to as the many-body localization~\cite{oganesyan2007localization,vznidarivc2008many,abanin2019colloquium}. This confirms our previous theoretical reasoning based on the analysis of the inverse participation ratio $\text{IPR}_0$.

\section{Thermalization after quench} \label{sec:quench}
\subsection{Thermalization of the time-dependent system occupancy}
Let us now consider a dynamical scenario in which the system-bath setup is initialized in an eigenstate of the Hamiltonian $\hat{H}$, and then the Hamiltonian is quenched in the moment $t=0$ to a new form $\hat{H}'$. Specifically,  we consider a quench of the system energy level from $\epsilon_0$ to $\epsilon_0'$, leaving the other parameters unchanged. Following Sec.~\ref{sec:mod}, we work within the Heisenberg picture, considering the evolution of the time-dependent observable $\hat{p}_k(t)$. Using Eq.~\eqref{ewop}, the evolution of the occupancy of level $k$ can be expressed as
\begin{align} \label{eq:popdyn}
\langle \hat{p}_k(t) \rangle= \sum_{l=0}^{K-1} |a_{kl}(t)|^2 \langle \hat{n}_l \rangle.
\end{align}
The equation is analogous to Eq.~\eqref{popstat} used in the time-independent case. Therefore, we may use the same arguments as in Secs.~\ref{subsec:stattheor} and~\ref{subsec:grand} to bound the indicator of eigenstate thermalization,
\begin{align}
	I_\text{mc}[\hat{p}_k(t)] \lessapprox \frac{1}{2}\sqrt{\text{IPR}_k(t)}, \quad I_\text{gc}[\hat{p}_k(t)] \leq \frac{1}{2}\sqrt{\text{IPR}_k(t)},
\end{align}
where 
\begin{align}
\text{IPR}_k(t) \equiv \sum_{l=0}^{K-1} |a_{kl}(t)|^4
\end{align}
is the time-dependent inverse participation ratio.

%%%%%%%%%%%%%%%%%%%%%%%%%%%%%%%%%%%%%%%%%%%%%%%%%%%%%%%%%%%%%%%%%%%%
\begin{figure}
	\centering
	\includegraphics[width=0.9\linewidth]{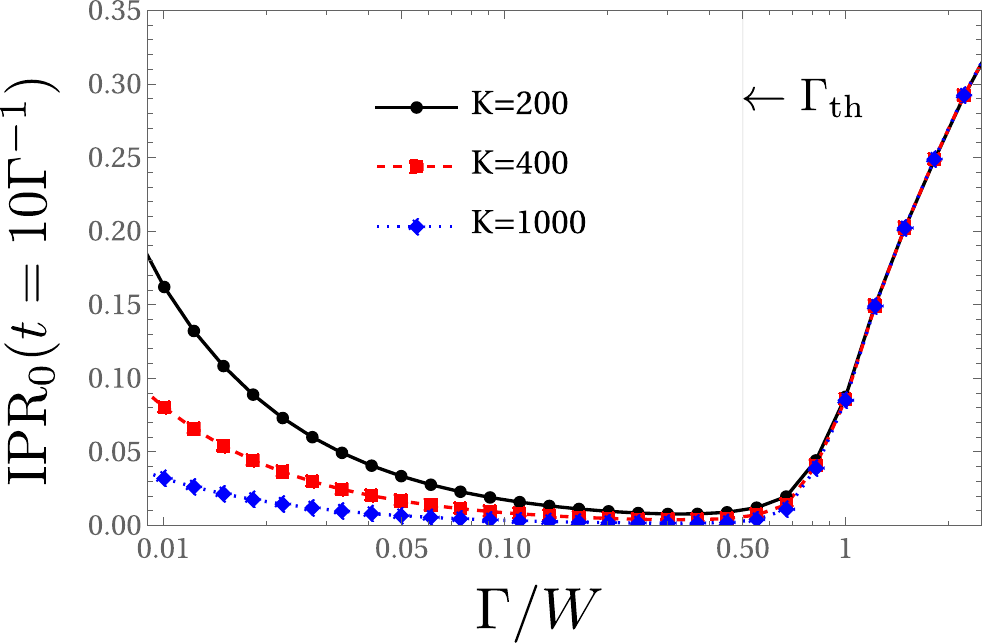}
	\caption{The time-dependent inverse participation ratio $\text{IPR}_0(t)$ as a function of the coupling strength $\Gamma$ for a fixed time $t=10 \Gamma^{-1}$ and different number of levels $K$. Parameters: $\epsilon_0=0.2W$, $\epsilon_0'=-0.2W$.}
	\label{fig:iprquench}
\end{figure}
%%%%%%%%%%%%%%%%%%%%%%%%%%%%%%%%%%%%%%%%%%%%%%%%%%%%%%%%%%%%%%%%%%%%
We now consider the case where the system energy is quenched from $\epsilon_0=0.2W$ to $\epsilon_0'=-0.2W$ at time $t=0$. In Fig.~\ref{fig:iprquench} we present the inverse participation ratio $\text{IPR}_0(t)$ as a function of $\Gamma$ for a fixed time $t=10 \Gamma^{-1}$. It exhibits a qualitatively similar behavior to the time-independent inverse participation ratio (corresponding to $t=0$) presented in Fig.~\ref{fig:ipr}. A notable difference is that the localization threshold $\Gamma_\text{th}$ is now shifted to a larger value of about $W/2$.

%%%%%%%%%%%%%%%%%%%%%%%%%%%%%%%%%%%%%%%%%%%%%%%%%%%%%%%%%%%%%%%%%%%%
\begin{figure}
	\centering
	\includegraphics[width=0.9\linewidth]{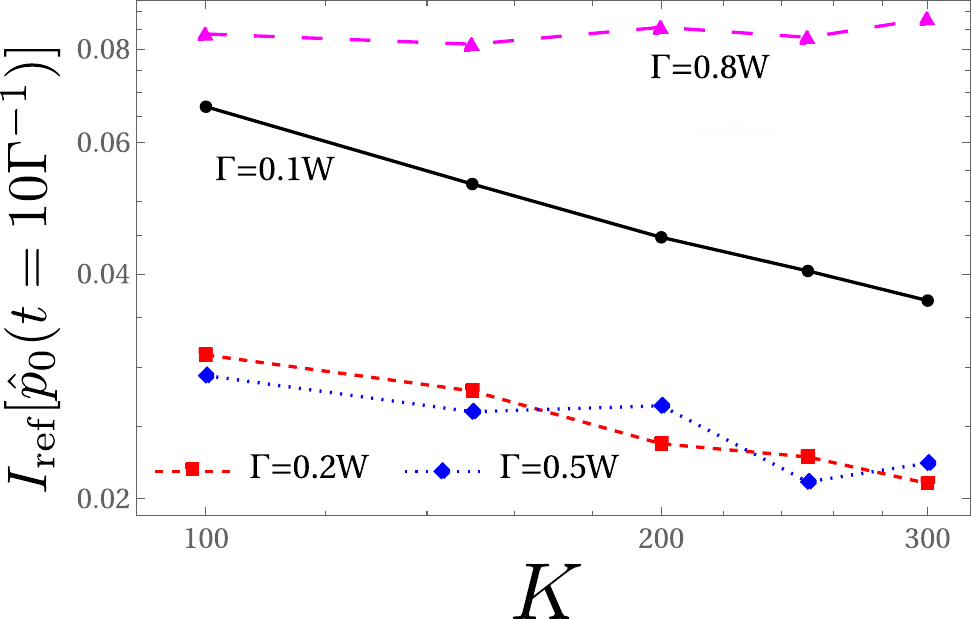}
	\caption{Scaling of the indicator of eigenstate thermalization $I_\text{ref}[\hat{p}_0(t)]$ with the number of levels $K$ for different values of the coupling strength $\Gamma$, evaluated for $M=100$ randomly generated eigenstates and a fixed time $t=10 \Gamma^{-1}$. The results are represented by dots, and lines are added for eye guidance. Parameters: $\epsilon_0=0.2 W$, $\epsilon_0'=-0.2W$.}
	\label{fig:popquenchscaling}
\end{figure}
%%%%%%%%%%%%%%%%%%%%%%%%%%%%%%%%%%%%%%%%%%%%%%%%%%%%%%%%%%%%%%%%%%%%

Following our discussion in Sec.~\ref{subsec:grand}, the presented scaling of $\text{IPR}_0(t)$ rigorously demonstrates thermalization of the system occupancy for typical eigenstates sampled from the grand canonical ensemble. For the case of sampling from the microcanonical ensemble, we apply the same procedure as in the previous section, with the same sample of $M=100$ randomly generated eigenstates.
In Fig.~\ref{fig:popquenchscaling} we present the scaling of the modified indicator of eigenstate thermalization $I_\text{ref} [\hat{p}_0(t)]$, defined as in Eq.~\eqref{refindicator}, for a fixed time $t=10 \Gamma^{-1}$. We can observe a decrease of the analyzed quantity with increasing $K$ for $\Gamma=0.1W$ and $\Gamma=0.2W$ belonging to the delocalized regime, which witnesses thermalization of the time-dependent system occupancy. Notably, a situation is now not fully clear for $\Gamma=0.5W$. This is because, as shown in Fig.~\ref{fig:iprquench}, this coupling value now corresponds to the border of the delocalized and localized phase, rather than being located within the localized phase. However, the effect of localization is clearly visible for an even larger $\Gamma=0.8 W$, for which the analyzed indicator does not vanish with $K$. 

\subsection{Thermalization of the system occupancy to the thermal value for the final Hamiltonian}

Thus far, we have shown that (in the delocalized regime and for large bath sizes) the evolution of the system occupancy is thermal in the sense that it tends to be the same independent of whether the system bath-setup is initialized in a typical eigenstate of its initial Hamiltonian or in a corresponding thermal state. Now, we go a step further by noting a well-established observation of open quantum system theory~\cite{jakvsic1996model, bach2000return, merkli2001positive, frohlich2004another, merkli2007decoherence, merkli2008resonance, farelly2017thermalization, dabelow2022thermalization, trushechkin2022open}: When the system-bath setup is initialized in the thermal state of the initial Hamiltonian $\hat{H}$, and then the Hamiltonian is weakly quenched, the reduced state of the system tends to relax to the equilibrium state for the new Hamiltonian. The justification for this occurrence and its conditions are thoroughly discussed in Ref.~\cite{trushechkin2022open}. In particular, the perturbation has to be sufficiently weak, e.g., affect only the system Hamiltonian, without changing the bath Hamiltonian (which is a scenario considered here). Thus, one may expect that the same relaxation to the equilibrium value for the new Hamiltonian would also be observed when the system-bath setup is initialized in a typical eigenstate of its initial Hamiltonian.

%%%%%%%%%%%%%%%%%%%%%%%%%%%%%%%%%%%%%%%%%%%%%%%%%%%%%%%%%%%%%%%%%%%%
\begin{figure}
	\centering
	\includegraphics[width=0.9\linewidth]{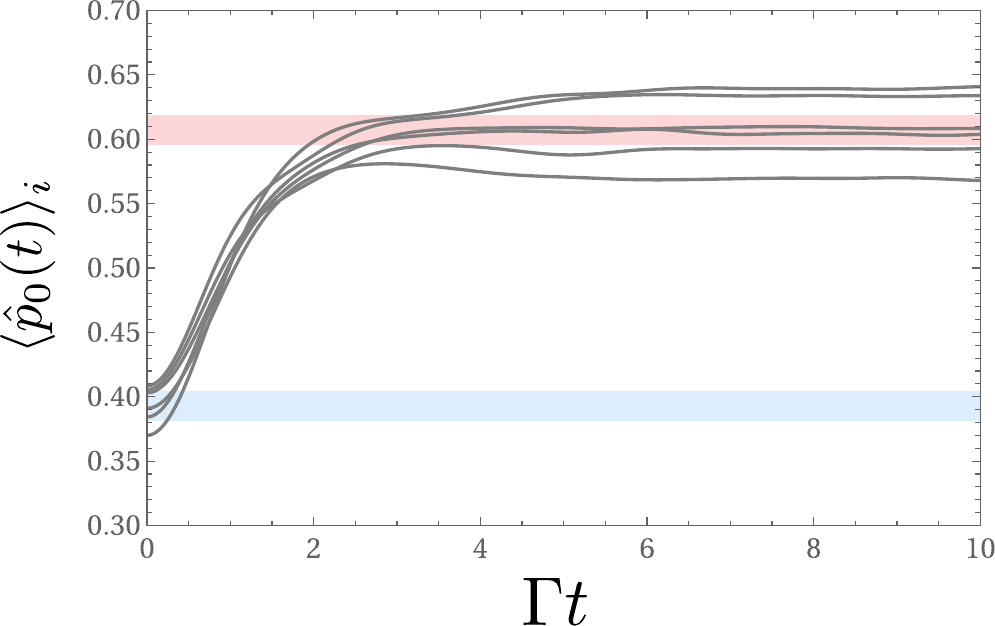}
	\caption{The evolution of the system occupancies $\langle \hat{p}_0 (t) \rangle_i$ for six individual random eigenstates $|E_i,N_i \rangle$ (gray solid lines). Blue- and red-shaded regions denote the range of allowed equilibrium occupancies of the system for the initial and the final Hamiltonian, respectively. Parameters: $\epsilon_0=0.2W$, $\epsilon_0'=-0.2W$, $\Gamma=0.2W$, $K=300$.}
	\label{fig:popquench}
\end{figure}
%%%%%%%%%%%%%%%%%%%%%%%%%%%%%%%%%%%%%%%%%%%%%%%%%%%%%%%%%%%%%%%%%%%%
To illustrate that, in Fig.~\ref{fig:popquench} we show the dynamics of the system occupancy $\langle \hat{p}_0(t) \rangle_i$ for a system-bath setup initialized in six different randomly generated eigenstates of the initial Hamiltonian.  We choose $\Gamma=0.2W$ corresponding to the delocalized regime. As in the static case, the shaded regions in the graph denote the range of allowed equilibrium occupancies of the system for the thermal states with the same mean particle number as the considered eigenstates and the temperatures belonging to the considered reference temperature window. The blue and red shading of the regions now correspond to allowed thermal values for the initial and final Hamiltonians. As follows from our previous analysis, the populations are initially distributed near the blue-shaded region, which witnesses thermalization of the initial conditions. Then, after the Hamiltonian quench, the populations evolve so that after a certain time of the order of $\Gamma^{-1}$ the populations focus around the red-shaded region. This suggests that the Hamiltonian quench indeed drives the system to thermalize with respect to the final Hamiltonian. (We note that, in contrast, the bath level occupancies do not thermalize, but rather remain close to their initial values throughout the system evolution.)

To demonstrate thermalization after quench more rigorously, let us further analyze  the behavior of time-averaged system occupancy $\overline{\langle \hat{p}_0(t)\rangle}$, where 
\begin{align}
\overline{(\cdot)} \equiv \lim_{\tau \rightarrow \infty} \frac{1}{\tau} \int_0^\tau (\cdot)dt
\end{align}
denotes the infinite-time average. Following Ref.~\cite{dalessio2016quantum}, thermalization after quench is defined as the fulfillment of two conditions:
\begin{itemize}
\item convergence of the time-averaged system occupancy $\overline{\langle \hat{p}_0(t) \rangle}$ to the thermal value for the new Hamiltonian;
\item vanishing of temporal fluctuations around the time-averaged value.
\end{itemize}
The fulfillment of the first condition can be witnessed by vanishing of the indicator of eigenstate thermalization defined analogously to Eq.~\eqref{refindicator},
\begin{align}
I_\text{ref}^\infty(\hat{p}_0) \equiv \sqrt{\frac{1}{M} \sum_{i=1}^M \left(\overline{\langle \hat{p}_0(t)\rangle_i}-\langle \hat{p}_0\rangle_{\text{gcf},i} \right)^2},
\end{align}
where $\overline{\langle \hat{p}_0(t)\rangle_i}$  is the time-averaged system occupancy for an individual eigenstate $i$, and $\langle \hat{p}_0\rangle_{\text{gcf},i}$
is the system occupancy for the thermal state of the final Hamiltonian with the reference temperature $T_i$ and chemical potential $\mu_i$ given by Eqs.~\eqref{eq:refen}--\eqref{eq:refnum}. These quantities can be calculated as~\cite{lydzba2023generalized}\footnote{More precisely, the former formula is valid provided that the energies of the normal modes of the final Hamiltonian $\nu_m$ are nondegenerate. This is fulfilled for the considered model with nonzero values of the tunnel couplings $t_k$~\cite{o1990computing}.}
\begin{align}
\overline{\langle \hat{p}_0(t) \rangle_i} &=\sum_{m=0}^{K-1}|b_{0m}|^2 \langle \hat{f}_m^\dagger \hat{f}_m \rangle_i, \\
\langle \hat{p}_0\rangle_{\text{gcf},i}  &=\sum_{m=0}^{K-1} |b_{0m}|^2 f[\beta_i (\nu_m-\mu_i)],
\end{align}
where the amplitudes $b_{0m}$ and the operators $\hat{f}_m$ were defined in Sec.~\ref{sec:mod}.

The fulfillment of the second condition is known as \textit{equilibration} and usually occurs under milder conditions than thermalization~\cite{gogolin2016equilibration}. It can be witnessed by the vanishing of the temporal standard deviation
\begin{align}
\sigma_i \equiv \sqrt{\overline{\langle \hat{p}_0(t) \rangle_i^2}-\overline{\langle \hat{p}_0(t) \rangle_i}^2},
\end{align}
which can be calculated as~\cite{lydzba2023generalized}
\begin{align}
\sigma_i=\sqrt{\sum_{l,m \neq l} |b_{0l}|^2 |b_{0m}|^2 |\langle \hat{f}_l^\dagger \hat{f}_m \rangle_i|^2}.
\end{align}
More precisely, we consider the averaged standard deviation
\begin{align}
\sigma_\text{av}=\frac{1}{M} \sum_{i=1}^M \sigma_{i}.
\end{align}

%%%%%%%%%%%%%%%%%%%%%%%%%%%%%%%%%%%%%%%%%%%%%%%%%%%%%%%%%%%%%%%%%%%%
\begin{figure}
	\centering
	\includegraphics[width=0.9\linewidth]{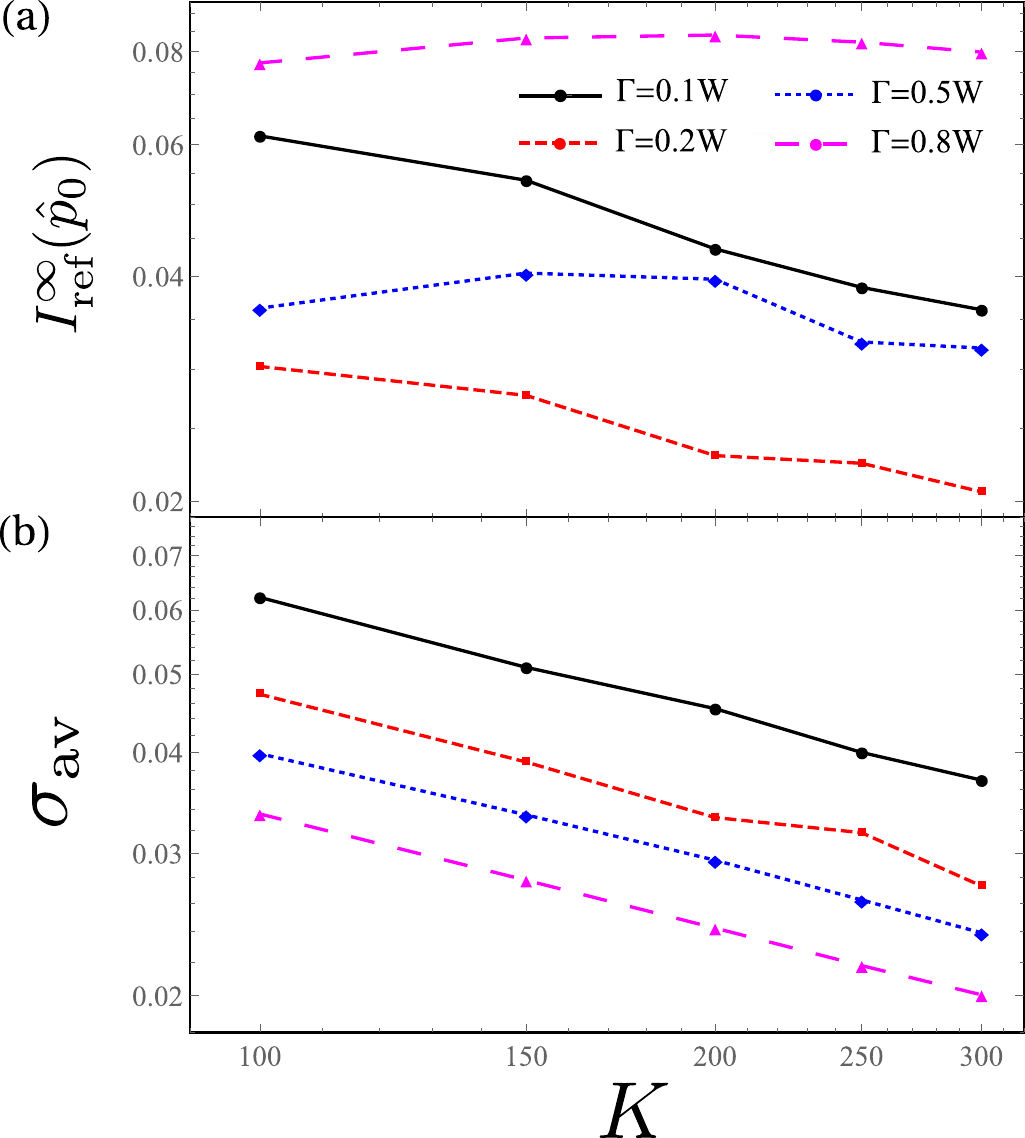}
	\caption{Scaling of the indicator of eigenstate thermalization $I^\infty_\text{ref}(\hat{p}_0)$ (a) and temporal fluctuations $\sigma_\text{av}$ (b) with the number of levels $K$. Other plot details as in Fig.~\ref{fig:popquenchscaling}.}
	\label{fig:time-average}
\end{figure}
%%%%%%%%%%%%%%%%%%%%%%%%%%%%%%%%%%%%%%%%%%%%%%%%%%%%%%%%%%%%%%%%%%%%

We plot $I_\text{ref}^\infty(\hat{p}_0)$ and $\sigma_\text{av}$ in Fig.~\ref{fig:time-average}. As shown, for small $\Gamma=0.1W$ and $\Gamma=0.2W$, corresponding to the delocalized regime, $I_\text{ref}^\infty(\hat{p}_0)$ decays with the setup size approximately as $1/\sqrt{K}$, witnessing the fulfillment of the first condition of thermalization. In contrast, it does not vanish for large $\Gamma=0.5W$ and $\Gamma=0.8W$, indicating the absence of thermalization in the localized regime. Furthermore, the temporal fluctuations $\sigma_\text{av}$ decay approximately as $1/\sqrt{K}$ independently of $\Gamma$, demonstrating the fulfillment of the second condition. (We note that while such a scaling can be proven for delocalized observables~\cite{lydzba2023generalized}, in our model the system occupancy appears to typically equilibrate also in the localized regime of large $\Gamma$.\footnote{We also note that the lack of equilibration after  quench was reported in a recent study~\cite{purkayastha2024difference} of a model closely related to the one analyzed in our work, where the system-bath setup was initialized in a typical pure state (superposition of many eigenstates) sampled from the grand canonical ensemble. However, that study was limited to much smaller bath sizes ($K-1 \leq 26$), which could have made the observation of equilibration difficult.}) This demonstrates that thermalization after quench occurs in the delocalized regime of small $\Gamma$.

The presence of thermalization in the considered model is not obvious, as integrable systems may not thermalize after quench, even when they exhibit weak eigenstate thermalization of static observables due to strong overlap of the initial state with rare nonthermal eigenstates of the final Hamiltonian, which can occur even for typical initial conditions~\cite{biroli2010effect, nandy2016eigenstate}. (However, thermalization after quench has been observed in certain regimes of integrable spinor condensates~\cite{dag2018classification}.) As already mentioned, the presence of thermalization in our model is related to the fact that we quench only the system Hamiltonian, leaving the bath Hamiltonian unchanged. Thus, the considered quench is only a small perturbation to the total system-bath Hamiltonian. Thermalization would not be observed for stronger perturbations, involving also the bath Hamiltonian. This does not invalidate the significance of our result, as quenches affecting only the system Hamiltonian are very relevant from the experimental point of view. For example, they correspond to typical experiments in the field of nanothermodynamics, performed on quantum dots or single-electron transistors attached to electrodes~\cite{koski2013distribution, koski2014experimental, maillet2019optimal, scandi2022minimally}.

\section{Thermalization induced by bath eigenstates} \label{sec:bath}
So far we have focused on the case where the system-bath setup was initialized in an eigenstate of the total Hamiltonian $\hat{H}$. Let us now consider another case, where the initial state of the system is arbitrary, while the bath is initialized in an eigenstate of its Hamiltonian
\begin{align}
\hat{H}_B =\sum_{k=1}^{K-1} \epsilon_k \hat{c}_k^\dagger \hat{c}_k.
\end{align}
Such eigenstates are then denoted as $|E^B_i,N^B_i \rangle$. The initial state of the system-bath setup reads then
\begin{align}
\rho_i(0)=\rho_S(0) \otimes |E^B_i,N^B_i \rangle \langle E^B_i,N^B_i |,
\end{align}
where $\rho_S(0)$ is the initial state of the system. A similar scenario, in which the baths were initialized in individual eigenstates sampled from the grand canonical ensemble, was recently investigated in Ref.~\cite{usui2023microscopic}.

Let us also define the initial state
\begin{align}
\rho_\text{mc}(0) \equiv \rho_S(0) \otimes \rho_\text{mc}^B,
\end{align}
which corresponds to the initial microcanonical state of the bath. Here $\rho_\text{mc}^B$ is defined via the right-hand-side of Eq.~\eqref{eq:microcan} with $|E_i,N_i \rangle$ replaced by $|E^B_i,N^B_i \rangle$. We now compare the dynamics of the system occupancy generated by single eigenstates and the microcanonical state. As previously, we work in the Heisenberg picture. The dynamics of the system occupancy is described via Eq.~\eqref{eq:popdyn}. The quench of the Hamiltonian now corresponds to the switching-on of the tunnel coupling $\Gamma$ at the time $t=0$. Since the system and the bath are initially uncoupled, we have $\hat{c}_k=\hat{d}_k$, $\hat{p}_k=\hat{n}_k$ and $a_{kl}=\delta_{kl}$.

%%%%%%%%%%%%%%%%%%%%%%%%%%%%%%%%%%%%%%%%%%%%%%%%%%%%%%%%%%%%%%%%%%%%
\begin{figure}
	\centering
	\includegraphics[width=0.9\linewidth]{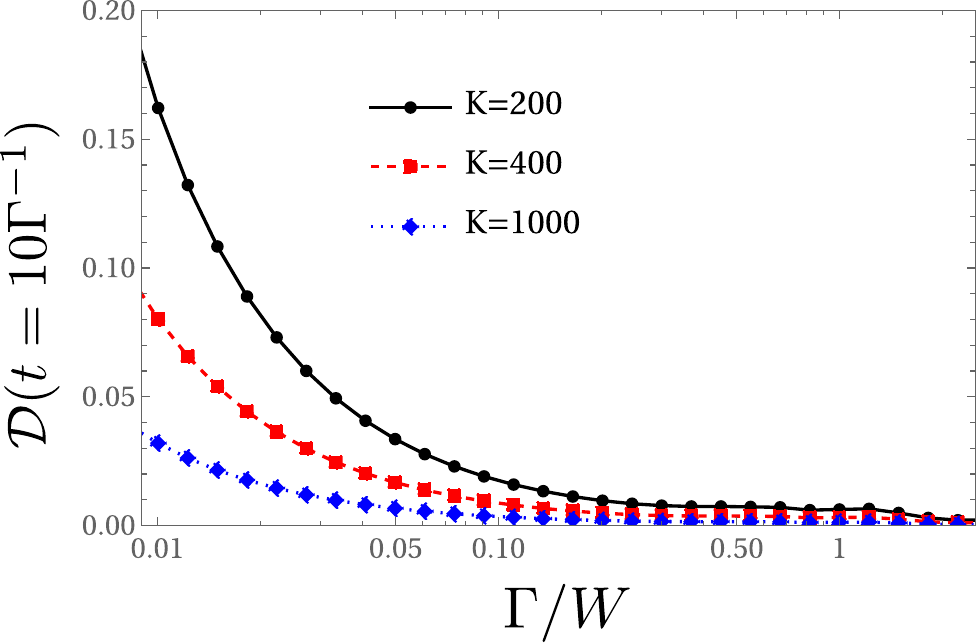}
	\caption{The localization coefficient $\mathcal{D}(t)$ as a function of the coupling strength $\Gamma$ for a fixed time $t=10 \Gamma^{-1}$ and different number of levels $K$. Parameters: $\epsilon_0=-0.2W$.}
	\label{fig:loccoef}
\end{figure}
%%%%%%%%%%%%%%%%%%%%%%%%%%%%%%%%%%%%%%%%%%%%%%%%%%%%%%%%%%%%%%%%%%%%
Let us now consider the indicator of eigenstate thermalization defined in Eq.~\eqref{indicator}, where we now take $\langle \hat{A} \rangle_i \equiv \Tr[\rho_i(0) \hat{A}]$ and $\langle \hat{A} \rangle_\text{mc} \equiv \Tr[\rho_\text{mc}(0) \hat{A}]$. Using the same arguments as before, it can be bounded as
\begin{align} \label{eq:boundbathmc}
	I_\text{mc}[\hat{p}_0(t)] \lessapprox \frac{1}{2}\sqrt{\mathcal{D}(t)},
\end{align}
where 
\begin{align}
\mathcal{D}(t) \equiv \text{IPR}_0(t)-|a_{00}(t)|^4=\sum_{l=1}^{K-1} |a_{0l}(t)|^4
\end{align}
is the localization coefficient. It corresponds to the previously defined time-dependent inverse participation ratio $\text{IPR}_0(t)$ with an excluded element $|a_{00}(t)|^4$; this is because the initial system occupancy is now the same for the states $\rho_i(0)$ and $\rho_\text{mc}(0)$. In Fig.~\ref{fig:loccoef} we plot this quantity as a function of $\Gamma$ for a fixed time $t=10 \Gamma^{-1}$. One may observe that, in contrast to the inverse participation ratio, the localization coefficient does not imply localization for a large $\Gamma$. Therefore, the thermal behavior of the system occupancy can be expected regardless of coupling strength $\Gamma$. We note that the formation of bound states for large $\Gamma$ still manifests itself in the nonvanishing long-time value of the element $|a_{00}(t)|^4$, which corresponds to the long-time memory of the system about its initial state. However, this element does not contribute to $\mathcal{D}(t)$. 

%%%%%%%%%%%%%%%%%%%%%%%%%%%%%%%%%%%%%%%%%%%%%%%%%%%%%%%%%%%%%%%%%%%%
\begin{figure}
	\centering
	\includegraphics[width=0.9\linewidth]{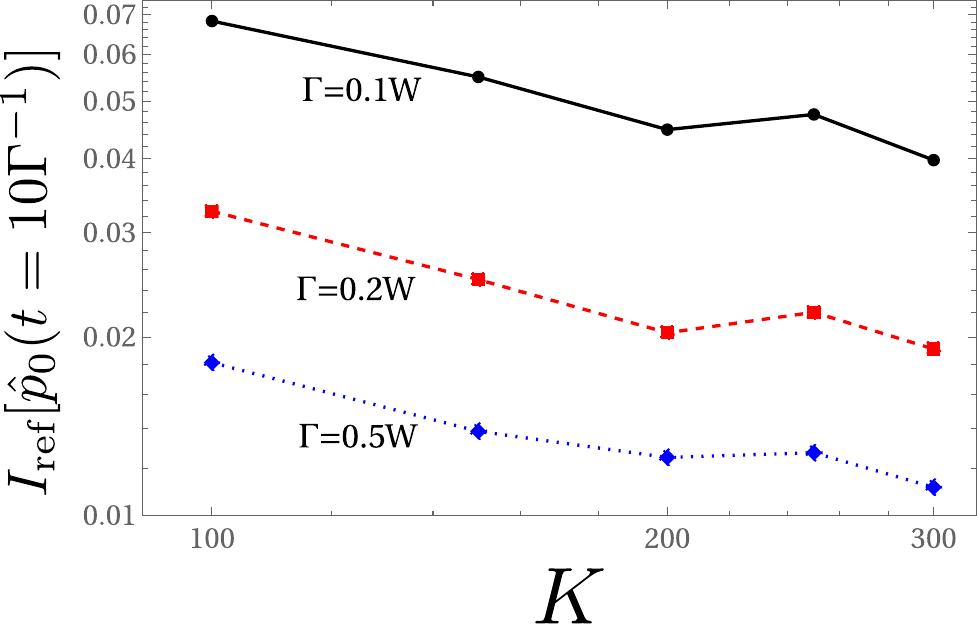}
	\caption{Scaling of the indicator of eigenstate thermalization $I_\text{ref}[\hat{p}_0(t)]$ with the number of levels $K$ for different values of the coupling strength $\Gamma$, evaluated for $M=100$ randomly generated bath eigenstates and a fixed time $t=10 \Gamma^{-1}$. The results are represented by dots, and lines are added for eye guidance. Parameters: $\langle \hat{p}_0(0) \rangle=0$, $\epsilon_0=-0.2 W$.}
	\label{fig:popbathscaling}
\end{figure}
%%%%%%%%%%%%%%%%%%%%%%%%%%%%%%%%%%%%%%%%%%%%%%%%%%%%%%%%%%%%%%%%%%%%
Let us now analyze the modified indicator of eigenstate thermalization defined via Eq.~\eqref{refindicator}. The system occupancy for the reference thermal state is now defined as $\langle \hat{p}_0(t) \rangle_{\text{gc},i} \equiv \Tr [\rho_{\text{gc},i}(0) \hat{p}_0(t)]$, where $\rho_{\text{gc},i}(0) \equiv \rho_S(0) \otimes \rho^B_{\text{gc},i}$ and $\rho^B_{\text{gc},i}$ is the reference grand canonical state of the bath with the same energy and particle number as the considered eigenstate. We further focus on the case where the system is initially empty $(\langle \hat{p}_0(0) \rangle=0)$ and the bath is initially half-filled [$N_i^B=(K-1)/2$]. As before, we consider only those eigenstates of the bath, whose reference temperatures $T_i$ belong to the interval $[T_\text{mid}-\Delta T/2,T_\text{mid}+\Delta T/2]$ with $T_\text{mid}=0.45W$ and $\Delta T=0.1W$. The scaling of the analyzed quantity with $K$ for different coupling strengths $\Gamma$ is presented in Fig.~\ref{fig:popbathscaling}; here we use the same random eigenstates for all values of $\Gamma$. As in the cases considered above, it decreases with the setup size (with certain statistical errors due to the finite sample size), implying thermalization in the thermodynamic limit $K \rightarrow \infty$. As suggested by Fig.~\ref{fig:loccoef}, this is also observed for large values of $\Gamma$, since the localization now does not suppress the thermalization.

%%%%%%%%%%%%%%%%%%%%%%%%%%%%%%%%%%%%%%%%%%%%%%%%%%%%%%%%%%%%%%%%%%%%
\begin{figure}
	\centering
	\includegraphics[width=0.9\linewidth]{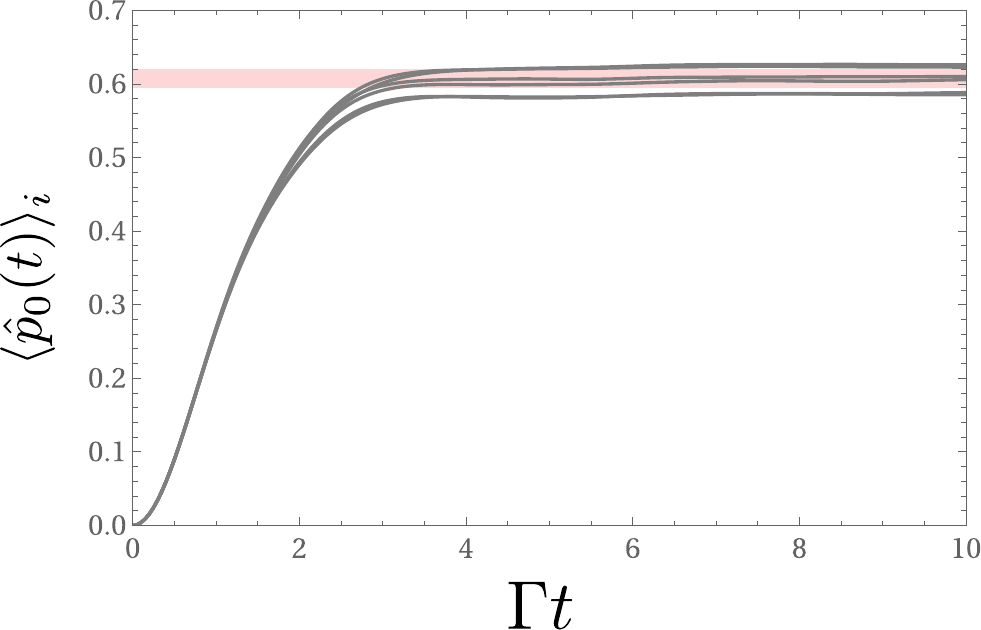}
	\caption{The evolution of the system occupancies $\langle \hat{p}_0 (t) \rangle_i$ for six individual initial random bath eigenstates $|E_i^B,N_i^B \rangle$ (gray solid lines) for an intermediate system-bath coupling $\Gamma=0.2W$. The red-shaded region denotes the range of allowed equilibrium occupancies. Parameters: $\langle \hat{p}_0(0) \rangle=0$, $\epsilon_0=-0.2W$, $K=301$.}
	\label{fig:popbath}
\end{figure}
%%%%%%%%%%%%%%%%%%%%%%%%%%%%%%%%%%%%%%%%%%%%%%%%%%%%%%%%%%%%%%%%%%%%
Finally, as in Sec.~\ref{sec:quench}, we analyze the evolution of the system occupancy to establish whether it relaxes to the equilibrium value for the total Hamiltonian. In Fig.~\ref{fig:popbath} we show dynamics of the system occupancy for the bath initialized in six different random eigenstates for an intermediate system-bath coupling $\Gamma=0.2W$. As before, the red-shaded region denotes the range of allowed equilibrium system occupancies for the chosen Hamiltonian parameters and the temperatures belonging to the considered reference temperature window. Analogously to the case of Hamiltonian quench, after a certain relaxation time, the system occupancies focus around the red-shaded region, which witnesses the system relaxation to equilibrium. We also note that, as in the previous section, the bath level occupancies do not thermalize but rather stay close to initial values.

%%%%%%%%%%%%%%%%%%%%%%%%%%%%%%%%%%%%%%%%%%%%%%%%%%%%%%%%%%%%%%%%%%%%
\begin{figure}
	\centering
	\includegraphics[width=0.9\linewidth]{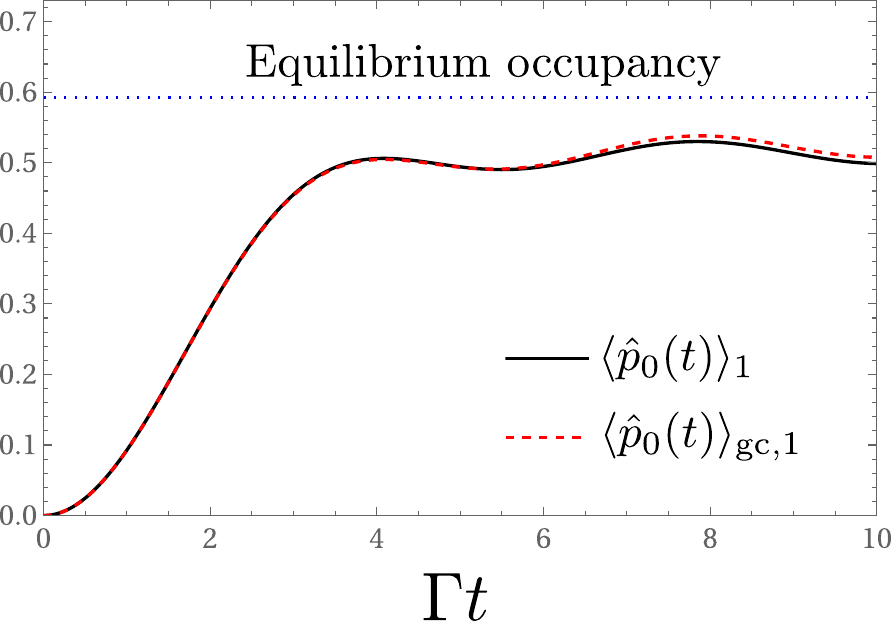}
	\caption{The evolution of the system occupancies for an initial bath eigenstate $|E_1^B,N_1^B \rangle$ (black solid line) and the reference grand canonical state (red solid line) for a strong system-bath coupling $\Gamma=0.8W$. The blue dotted line denotes the equilibrium occupancy evaluated for the reference temperature $T_1$ and the chemical potential $\mu_1$. Other parameters as in Fig.~\ref{fig:popbath}.}
	\label{fig:popbath-bound}
\end{figure}
%%%%%%%%%%%%%%%%%%%%%%%%%%%%%%%%%%%%%%%%%%%%%%%%%%%%%%%%%%%%%%%%%%%%
In contrast, as in Sec.~\ref{sec:quench}, for a strong system-bath coupling, the system occupancy does not relax to the thermal value for the total Hamiltonian. To illustrate that, in Fig.~\ref{fig:popbath-bound} we present the evolution of the system occupancy for a large coupling strength $\Gamma=0.8W$. We focus on the evolution for a single initial state $\rho_1(0)$ (black solid line). It is compared with the results for a reference grand canonical state (red dashed line). The blue dotted line denotes the equilibrium occupancy calculated for the temperature $T_1$ and the chemical potential $\mu_1=0$ of the reference grand canonical state. We can see that now the system occupancy does not converge to its equilibrium value for either the initial eigenstate or the initial reference thermal state. This is characteristic for systems with bound states due to the long-time memory of the system about its initial state, which is encoded in the amplitude $|a_{00}(t)|^2$~\cite{cai2014threshold, xiong2015non, yang2015master, jussiau2019signature}. However, as indicated by the vanishing of  $I_\text{ref}[\hat{p}_0(t)]$, the system behavior is still thermal in the sense that the evolution of the system occupancy induced by the initial eigenstate of the bath and the reference thermal state is approximately the same.

\textit{Thermalization induced by typical bath eigenstates sampled from the grand canonical ensemble.---}We finally recall that Ref.~\cite{usui2023microscopic} recently numerically demonstrated thermalization in a similar scenario, but with the bath initialized in individual eigenstates randomly sampled from the grand canonical (rather than microcanonical) ensemble. Such thermalization can be witnessed by the vanishing of the corresponding indicator of eigenstate thermalization 
\begin{align} \label{indicatorgrand-bath}
	I_\text{gc}[\hat{p}_0(t)] \equiv \sqrt{\sum_{i} \frac{e^{-\beta(E_i^B-\mu N_i^B)}}{Z^B}  (\langle \hat{p}_0(t) \rangle_i-\langle \hat{p}_0(t) \rangle_{\text{gc}})^2},
\end{align}
where $Z_B=\sum_i e^{-\beta(E_i^B-\mu N_i^B)}$, $\langle \hat{A} \rangle_{\text{gc}} \equiv \Tr[\rho_\text{gc}(0) \hat{A}]$, $\rho_\text{gc}(0) \equiv \rho_S(0) \otimes \rho_\text{gc}^B$, and 
\begin{align}
\rho_\text{gc}^B=Z_B^{-1} \sum_i e^{-\beta(E_i^B-\mu N_i^B)} |E^B_i,N^B_i \rangle \langle E^B_i,N^B_i|   
\end{align} 
is the grand canonical state of the bath. Analogously to Eq.~\eqref{indicatorgrand}, it quantifies the typical deviation of the time-dependent system occupancies  $\langle \hat{p}_0(t) \rangle_i$, obtained for the bath initialized in the individual eigenstates, from the occupancy $\langle \hat{p}_0(t) \rangle_{\text{gc}}$ obtained for the initial grand canonical state of the bath. Applying the results of Sec.~\ref{subsec:grand}, this indicator obeys an exact bound similar to Eq.~\eqref{eq:boundbathmc},
\begin{align}
	I_\text{gc}[\hat{p}_0(t)] \leq \frac{1}{2}\sqrt{\mathcal{D}(t)}.
\end{align}
Thus, vanishing of $\mathcal{D}(t)$ (demonstrated in Fig.~\ref{fig:loccoef}) leads to thermalization in the discussed sense, which provides a theoretical justification for the observations of Ref.~\cite{usui2023microscopic}.

\section{Conclusions} \label{sec:concl}
We investigated the behavior of observables for individual eigenstates of a fully integrable noninteracting resonant level model. First, we have shown that typical eigenstates exhibit thermalization of the system occupancy, that is, the occupancy tends to have the same value as for the thermal state with the same energy and particle number as the considered eigenstate. Such thermalization for typical eigenstates is known in the literature as weak eigenstate thermalization. However, in contrast to previously demonstrated examples of this phenomenon, our model is neither translationally invariant~\cite{iyoda2017fluctuation,biroli2010effect,nandy2016eigenstate,lai2015entanglement,li2016quantum,dag2018classification}, random~\cite{magan2016random,cattaneo2024thermalization}, nor chaotic~\cite{lydzba2024normal}. In our case, thermalization is related to delocalization of the single-particle state, corresponding to the occupied state of the system, over many single-particle eigenstates of the system-bath Hamiltonian. As a consequence, the system occupancy can be expressed as a weighted average of occupancies of the normal modes. As shown by Khinchin in the context of classical systems~\cite{khinchin1949mathematical}, such averages of many independent degrees of freedom tend to thermalize in the thermodynamic limit due to the law of large numbers, independent of details of the system dynamics. We further show that the system thermalization becomes suppressed when the system occupancy becomes localized in a few normal modes due to formation of the bound states for a strong system-bath coupling. Furthermore, thermalization is not observed for the occupancies of the bath levels, which are always strongly localized.

We further went beyond the static scenario to consider the case where the system Hamiltonian is quenched at some moment of time. We have shown that after such a quench, the system occupancy tends to relax to the thermal value corresponding to a new Hamiltonian. This is nontrivial, as integrable systems may not thermalize after quench, even when they exhibit weak eigenstate thermalization of static observables~\cite{biroli2010effect, nandy2016eigenstate} (however, see Ref.~\cite{dag2018classification} for the observation of thermalization in integrable spinor condensates). The presence of thermalization in our model is related to the fact that the quench involves only parameters of the system and not of the bath. Therefore, it only weakly perturbs the total system-bath Hamiltonian. Such quenches that involve only the system Hamiltonian are common in experiments on open quantum systems~\cite{koski2013distribution, koski2014experimental, maillet2019optimal, scandi2022minimally}.

Finally, we considered the case where the initial state of the system is arbitrary, while the bath is initialized in an eigenstate of its Hamiltonian. Then, for typical eigenstates, we observed the same thermalization behavior of the system as for baths initialized in thermal states with the same energy and particle number. While we focused on a system coupled to a single bath, this approach can easily be generalized to a multiple-bath scenario. In such a case, we may expect the emergence of nonequilibrium steady states, as previously observed for baths obeying ETH~\cite{xu2022typicality, xu2023emergence,zhang2024emergence} or initialized in typical eigenstates sampled from the grand canonical ensemble~\cite{usui2023microscopic}.

We emphasize that the observed thermalization for typical eigenstates is not a direct consequence of the previous results on thermalization for typical (Haar-random) pure states with a narrow energy distribution~\cite{tasaki1998quantum, gemmer2003distribution, goldstein2006canonical, popescu2006entanglement} (i.e., typical coherent superpositions of eigenstates from the microcanonical shell) or with fixed average energy~\cite{muller2011concentration}. In those cases, the normal mode occupancies $\langle \hat{n}_l \rangle$ thermalize themselves, and thus all level occupancies $\langle \hat{p}_k \rangle$ (including the occupancies of the bath levels) also thermalize by virtue of Eq.~\eqref{popstat}. In our case, for individual eigenstates, the occupancies of normal modes are far from being thermal (they are always equal to 0 or 1), and system thermalization occurs only due to the aforementioned delocalization effects, leading to the averaging over many normal modes.

The paper contributes to the ongoing debate on whether nonintegrability should be regarded as a requirement for thermalization~\cite{de2015necessity, bartsch2017necessity, rigol2012alternatives, chakraborti2022entropy,baldovin2021statistical, baldovin2023ergodic, cocciaglia2022thermalization, cattaneo2024thermalization}. It is known that integrable systems do not exhibit thermalization of \textit{all} observables in \textit{all} physical scenarios~\cite{biroli2010effect, nandy2016eigenstate, li2016quantum, rigol2007relaxation, cassidy2011generalized}. In the considered system, this is illustrated by the lack of thermalization of bath level occupancies, or suppression of the system thermalization in the localized regime. The system will also not thermalize after the quench of the bath Hamiltonian. Nevertheless, the paper supports the conclusion of Refs.~\cite{baldovin2021statistical, chakraborti2022entropy, usui2023microscopic, baldovin2023ergodic, cocciaglia2022thermalization, cattaneo2024thermalization} that integrable systems can still exhibit a genuine thermal behavior of physically meaningful quantities in many physically relevant scenarios. Therefore, nonintegrability and ETH should not be regarded (as sometimes done~\cite{de2015necessity, deutsch2018eigenstate}) as the sole mechanism to explain the origin of thermalization. Rather, there is value in using different complementary approaches (including arguments based on the typicality of states and observables) to get a complete picture of the emergence of thermodynamic behavior.

\acknowledgments

We thank Peter Reimann for useful comments and references on thermalization after quench in integrable models.

\bibliography{bibliography}	
	
\end{document}